\pdfoutput=1
\documentclass[10pt]{IEEEtran}
\IEEEoverridecommandlockouts
\usepackage[utf8]{inputenc}
\usepackage[english]{babel}
\usepackage{graphicx}
 \usepackage{graphics}
\usepackage{epsfig}
\usepackage{amsmath,amssymb,amsthm}
\usepackage{amsthm,amsmath,amsfonts,ed,empheq}
\usepackage{color}
\usepackage{amsfonts}
\usepackage[lofdepth,lotdepth]{subfig}
\usepackage{mathrsfs}
\usepackage{algorithm}
\usepackage{algorithmic}
\usepackage{caption}
\usepackage{multicol}
\usepackage{lipsum}
\usepackage[numbers,sort&compress]{natbib}
\usepackage{multirow}
\usepackage{setspace}
\usepackage{xcolor,mdframed}

\def\Comment#1{\textit{#1}}

\newtheorem{proposition}{Proposition}

\newtheorem{definition}{Definition}
\newtheorem{remark}{Remark}

\begin{document}
\captionsetup[figure]{labelformat={default},labelsep=period,name={Fig.}}
\pagenumbering{gobble}


\title{Accurate Graph Filtering   in   Wireless Sensor Networks} 
\author{Leila Ben Saad,~\IEEEmembership{Member,~IEEE,} 
and Baltasar Beferull-Lozano,~\IEEEmembership{Senior Member,~IEEE}
\thanks{This work was supported in part by the PETROMAKS Smart-Rig grant
244205/E30, 
in part by the IKTPLUSS INDURB grant 270730/O70 and
in part by the TOPPFORSK WISECART grant 250910/F20 
from the Research Council of Norway.
This work was in part presented at SPAWC 2018 \cite{BenSaadLozano2018}.} 
\thanks{The authors are with Department of Information and Communication Technology, University
of Agder, Norway   (e-mails: leila.bensaad@uia.no, baltasar.beferull@uia.no). 
}
}

\markboth{Submitted to IEEE INTERNET OF THINGS JOURNAL}%
{Shell \MakeLowercase{\textit{et al.}}: Bare Demo of IEEEtran.cls for IEEE Journals}
\maketitle

\begin{abstract}


Wireless sensor networks (WSNs) are considered as a major technology enabling the Internet of Things (IoT) paradigm.
The  recent emerging Graph Signal Processing  field  can also contribute to enabling the IoT  by providing key tools, 
such as
graph filters, for  processing  the data  associated with the sensor devices.
 Graph filters can be  performed over WSNs in a distributed manner
 by means of a certain number of communication exchanges among the
nodes. But, WSNs  are often affected by 
 interferences and  noise, which leads to view these networks  as  directed, random and time-varying graph topologies.
Most of existing works neglect this problem by considering  an unrealistic assumption that
claims the same probability of link activation in both directions when sending a packet
between two neighboring nodes.
This work focuses on the problem of operating   graph filtering in random asymmetric WSNs. 
We  show first that  graph filtering with finite impulse response graph filters (node-invariant and node-variant)
requires having equal connectivity probabilities for all the links in order 
to have an unbiased filtering,  which {cannot} be achieved in practice in random WSNs.
After this, we characterize the graph  filtering
error and  present an efficient strategy  to conduct  graph
filtering tasks over random WSNs with  node-variant graph filters
by maximizing 
accuracy, that is, ensuring a small  bias-variance tradeoff.
In order to enforce the desired accuracy, we optimize the filter coefficients and
design a cross-layer distributed scheduling
algorithm at the MAC layer.
 Extensive numerical experiments are presented to show the efficiency of the proposed solution as well
 as the cross-layer distributed scheduling
algorithm for the denoising application.

\end{abstract}



\vspace{-0.2cm}
\section{Introduction}

\IEEEPARstart{T}{HE}  Internet  of  Things  (IoT) can be seen  as a technology that enables to connect a high
number of objects or ``Things" 
for the purpose
of exchanging information, and  where  Wireless Sensor Networks (WSNs) play an important role.
It is expected that these objects equipped with sensors can communicate with each other, perform in-network data processing  and  even take decisions by themselves.
The data collected by these objects often present  irregular and  complex structure that can no longer be processed by standard tools.
This has led recently to the emergence  of the Graph Signal Processing (GSP) field \cite{Shuman2013,Sandryhaila2014}, where  concepts and  tools 
such  as graph filters are used to analyze signals (i.e., sensor data) defined over a graph.
In this field,  graph filters (GFs) have been adopted in many tasks \cite{ShumanVandergheynstFrossard11,SandryhailaMoura2014BigData}
and used to solve several problems such as distributed estimation and consensus \cite{SandryhailaKar2014}, denoising and smoothing \cite{ChenSandryhaila14, MaHuangSegarra2016, IsufiLoukasSimonetto2017},  reconstruction \cite{NarangGadde13, GiraultGoncalves2014},
 and clustering \cite{TremblayPuy2016}.

 \begin{figure}[t]
 \vspace{-0.3cm}
\begin{center}
\subfloat[Original signal ]{\includegraphics[scale=0.35]{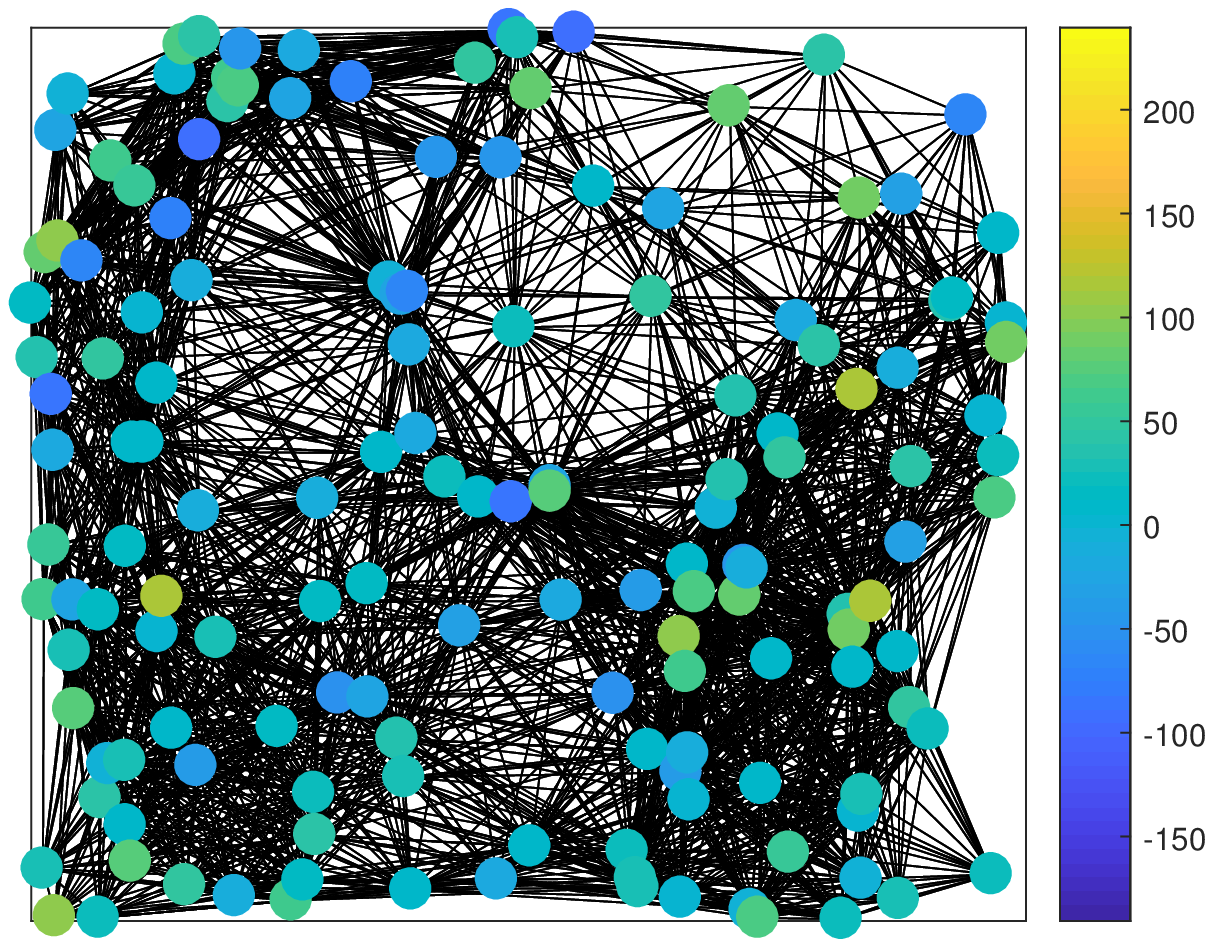}}
\subfloat[Original signal $+$ Noise  ]{\includegraphics[scale=0.35]{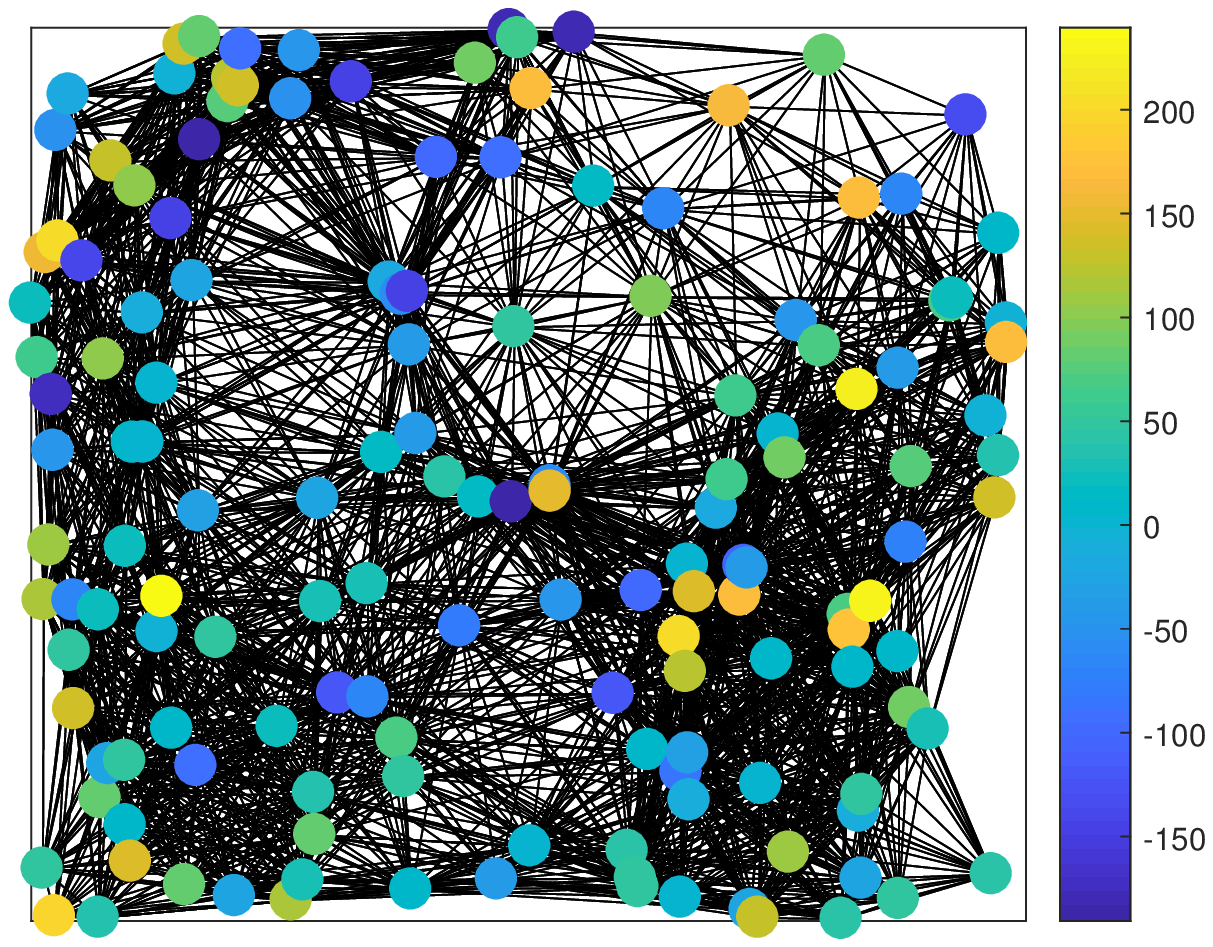}}\\
\vspace{-0.2cm}
\subfloat[Traditional protocol (RLBA) \cite{YuHua2012}]{\includegraphics[scale=0.35]{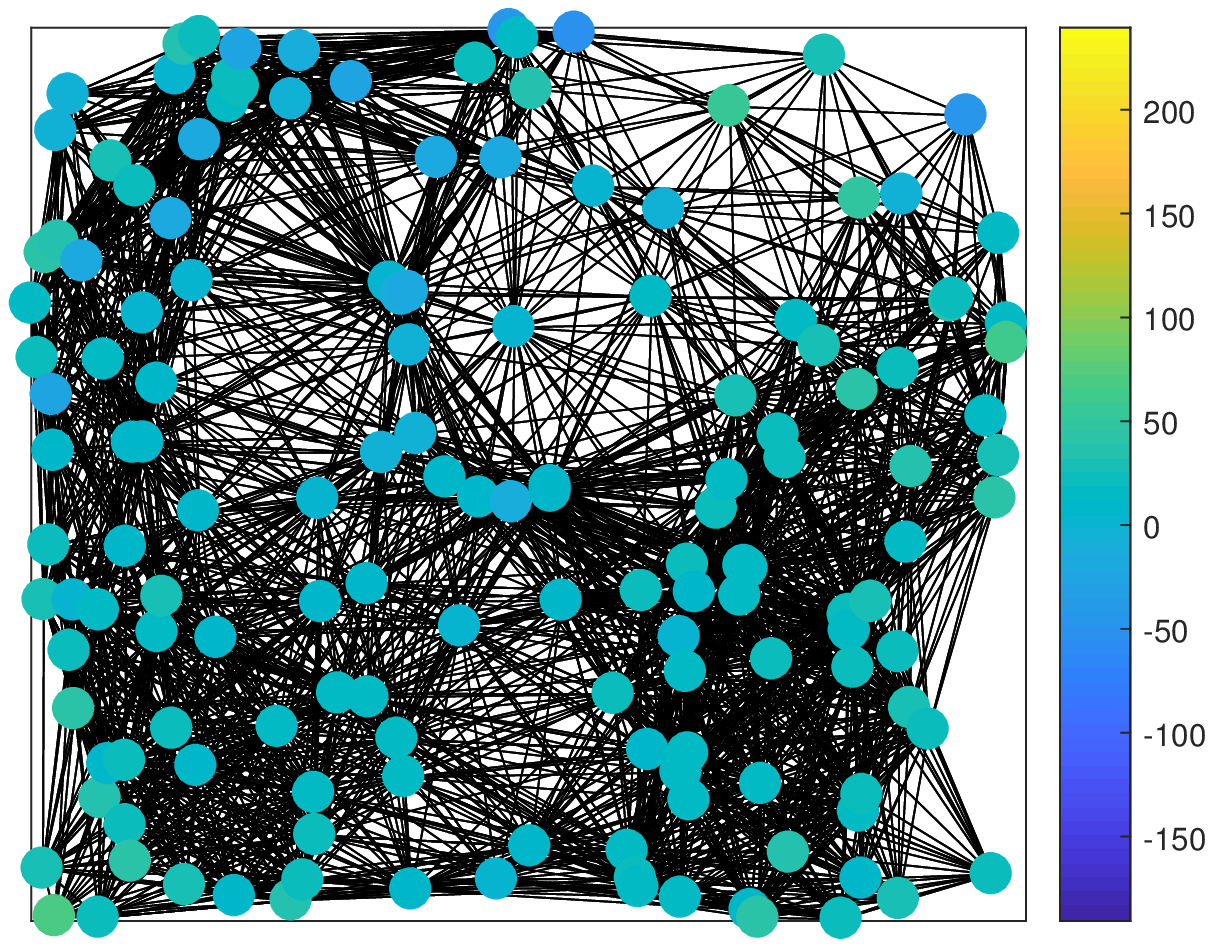}}
\subfloat[Cross-layer protocol]{\includegraphics[scale=0.35]{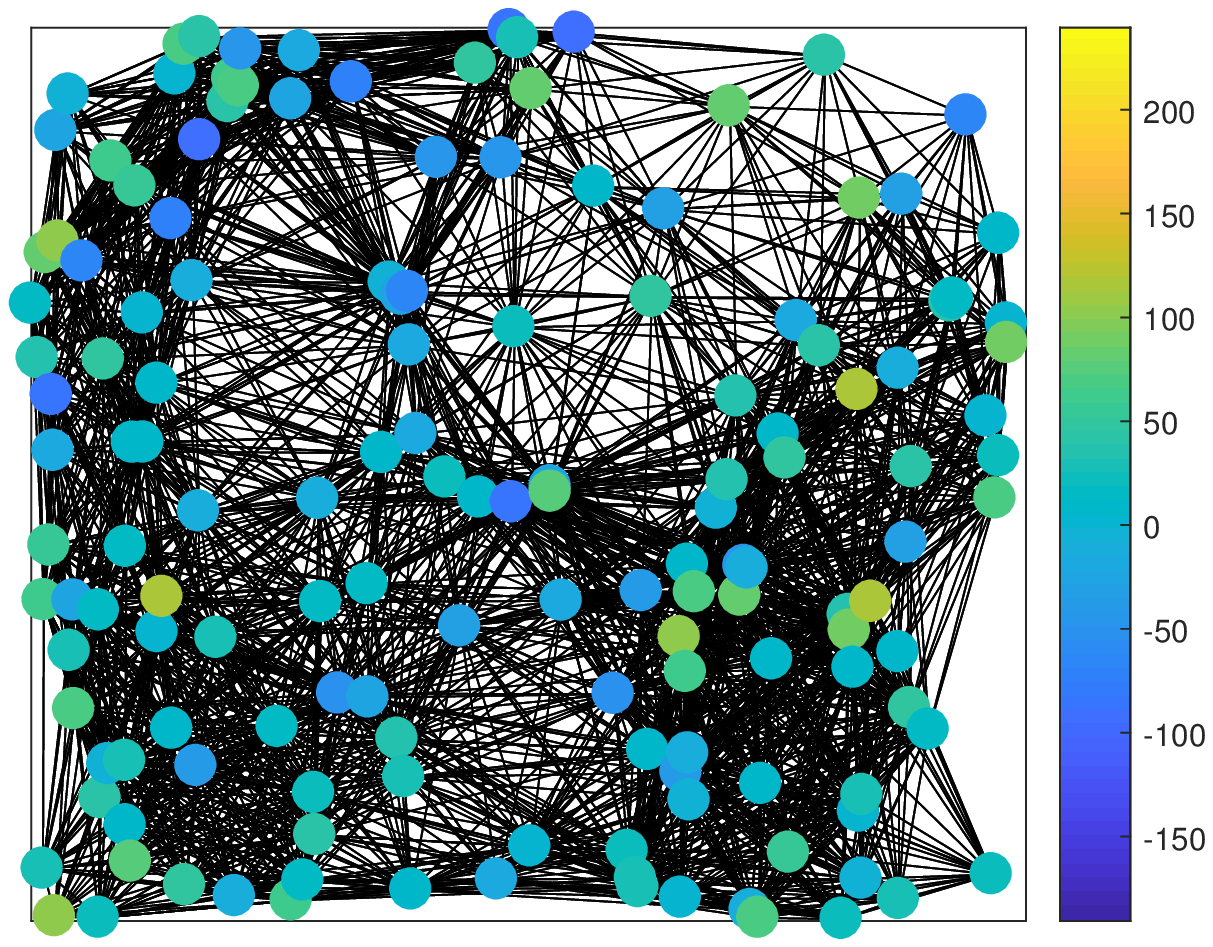}\label{fig-intro-CDSA}}
\end{center}
\caption{Denoising by graph filtering in a random wireless sensor network: designed protocol vs. traditional protocol. 
{Each circle represents a sensor node, where its  color depends on the signal value at that node. 
Different colors  mean higher difference between signal values at those nodes.}}
\label{fig-intro}
\vspace{-0.5cm}
\end{figure}

Due to their distributed implementations \cite{SegarraMarques17,IsufiLoukasSimonetto2017,IsufiLoukasSimonettoLeus17,ShumanVandergheynstFrossard11,BenSaadAsensio2017,CoutinoIsufiLeus2018},
graph filters can be  implemented distributedly
over WSNs \cite{NedicOlshevsky2018}.
However, it is noteworthy to emphasize on the fact that in typical deployments,
WSNs suffer from random and asymmetric 
packet losses, implying  to  view these networks  as directed, random   and  time-varying  graph topologies \cite{LifengAnish2010}. 
However, most of the works related to GFs do not consider the  problem of time-variability and randomness of the graph   when performing graph filtering tasks. 
In addition to that, the few works \cite{IsufiLoukasSimonetto2017,IsufiLoukasSimonettoLeus17,GamaIsufiLeus2018} that have analyzed this issue,
typically state that, when sending a packet
between two neighboring nodes,  the probability of link activation   is either  the same  in both directions or  equal for all
the links in the network. 
These unrealistic assumptions 
{cannot} be adopted in  real  conditions in WSNs, because the transmissions between nodes are often corrupted by
interferences and  noise \cite{ZamalloaKrishnamachari2007,LifengAnish2010}, creating asymmetry in the links.

In addition to solving the problem of link
asymmetry when performing graph filtering in WSNs, the
communication at the MAC layer should  be also considered because
to accomplish graph filtering tasks, the nodes need to
exchange data and therefore perform a high number of one-hop
transmissions. Broadcasting can be adopted in this case by activating simultaneously
several links  to reduce the delay of the execution of graph filtering tasks.
{There exist many distributed broadcast
scheduling algorithms \cite{GoussevskaiaMoscibroda2008,YuWang2011,YuHua2012,Halldorsson2012,DerbelTalbi2010,JurdzinskiKowalski2012,FuchsWagner2014,FuchsPrutkin2015}  that have been proposed in the
literature using the physical interference model \cite{GuptaKumar2000}, which reflects
more accurately the wireless medium. Among these algorithms,
the three algorithms proposed in \cite{GoussevskaiaMoscibroda2008,YuHua2012,FuchsPrutkin2015},
can be considered as among the most efficient existing algorithms
in terms of time complexity and fast medium access.}
Even though existing protocols can achieve successful local
broadcast with a high probability within a small number of
time slots, they are totally decoupled from the higher layers
and therefore, they are not designed to match the specific distributed computation needs
to ensure accurate graph filtering tasks (see Fig.~\ref{fig-intro}).

In this paper, we  first introduce the challenge of processing  graph filtering tasks with finite impulse response graph filters (node-invariant and node-variant) and show
that equal  probabilities for all the link connections enables to obtain an unbiased filtering (i.e., the expected output obtained in random time-varying graphs
is the same as for the underlying deterministic  graph without interference). 
Motivated by the fact that this  {cannot} be achieved in practice in realistic WSNs,
 then, we propose 
 to perform  graph
filtering  over random WSNs with  node-variant graph filters,
while providing an optimization  problem that finds the
filter coefficients as a bias-variance trade-off, allowing  to obtain
an accurate filtering over time-varying graphs.  
To enforce the accuracy of the graph filtering,
 which is    implemented distributedly by means 
of a certain number of communication exchanges among the nodes, we design  
a Cross-layer Distributed Scheduling Algorithm (CDSA) that controls the Packet Delivery Ratios (PDRs) at the nodes
such that this filtering accuracy is maximized, as illustrated  in  Fig.\ref{fig-intro}(d).

 The  main contributions of this paper can be summarized as follows:
 \begin{itemize}
    \item We 
    analyze   graph filtering with finite impulse response graph filters (node-invariant and node-variant)
and show that 
having equal connectivity probabilities for all the links enables to reach an unbiased filtering,  which {cannot} be achieved in practice in WSNs.
 \item {  We  characterize the graph  filtering
error and show the need of equalizing the probabilities of link
connections within the neighborhood of each transmitter 
in order to reduce the bias (expected error). This approach implies that
  the link activations (towards neighbors within a given transmission range) of a node are equal, but the link activations   across the different  broadcast 
regions corresponding to the different nodes are not.}
    \item We show how to conduct  graph
filtering tasks in random WSNs with  node-variant graph filters ensuring a high filtering accuracy,
by optimizing the filter coefficients that minimize a bias-variance trade-off. 

\item We propose a cross-layer distributed scheduling
algorithm that enables applying graph filters in WSNs under
asymmetric wireless links, while still achieving a  high  filtering accuracy. We  propose also to maximize 
the time efficiency of the filtering process by minimizing the total number of slots of the proposed protocol.

\item  We show through   numerical experiments that a  small normalized  squared error is
obtained when using our proposed   protocol  and
the optimized filtering coefficients,
  achieving  a good  performance for  the denoising application, as compared to using other state-of-art methods.
 \end{itemize}

The rest of this paper is structured as follows. Section~II presents the related work.
Section~III introduces
 the  main concepts related  to graph signal processing. 
 Section~IV analyzes  the challenge of  conducting graph filtering tasks in  
  random   WSNs.  In Section~V, 
 a  solution to overcome  this issue is proposed in order to ensure a high graph
filtering accuracy. 
Section~VI shows how to enforce this
accuracy at the MAC layer  by  designing  a  cross-layer distributed scheduling algorithm.
  Section~VII validates  our  results by experiments and 
Section~VIII presents the concluding remarks.


%
%
%

\textit{Notation and terminology}:
We indicate  vectors and  matrices by
bold lowercase letters and  uppercase letters, respectively. We represent the $(i,j)$th entry of a matrix $\bold A$  by $a_{ij}$.
The 2-norm of a vector $\bold u$ is denoted by $\| \bold u    \|$. We indicate   the spectral norm and the Frobenius norm of a matrix $\bold M$, respectively, by  $\| \bold M    \|_2$ and  $\| \bold M    \|_F$.
The notation  $\text{diag}(.)$, $\text{tr}(.)$  and  $\circ$ stands respectively for  the diagonal matrix, the trace operator and the Hadamard product.
We compute the covariance matrix  as
$\bold \Sigma_{\bold u} [t ] = \mathbb{E} [\bold u_t \bold u^H_t] - \mathbb{E} [\bold u_t]\mathbb{E} [\bold u_t]^H$, where
$\bold u_t$ is  a random process  at time $t$.

 \vspace{-0.15cm}
 \section{Related work}


Graph signal processing tasks, implemented in a distributed manner over random
WSNs, can be successfully accomplished 
if
they are  processed in an efficient manner by  other layers of the sensor nodes. At
the MAC layer, 
two classes of protocols \cite{HuangXiao2013} can be adopted: contention-free (scheduling) and
contention-based protocols.
Contention-free protocols avoid collisions between nodes by both dividing the
medium into a certain number of time slots and assigning each slot to one or multiple
 feasible nodes or links. Contrarily, contention-based protocols
allow the nodes to compete for medium access in a random and  asynchronous
  manner, but collisions {cannot} be completely
prevented. 
%
 When adopting   contention-free protocols 
to perform  graph signal processing tasks by means of communication exchanges between  sensor nodes, it is important 
 to consider  the interference problem.
 Different interference models have been adopted in the literature. 
 The most widely used 
 are the so-called \textit{protocol}  and   \textit{physical} models
 \cite{GuptaKumar2000}. In the protocol model, a communication
 between a transmitter node and a receiver node is successful if there is no
 other  node transmitting at the same time within a certain transmission range.
 In the physical interference model, a communication
 between a transmitter node and a receiver node is successful if the
 Signal to
Interference and Noise Ratio (SINR) at
the receiver is above a
certain threshold, whose value depends on the  channel characteristics.
 This means that in this model, the interference experienced by the receiver is not only caused
 by its neighbors inside its radio range but also by the nodes, which are further away.
For this reason,
the physical interference model reflects more accurately the wireless medium.
This  model is mainly suited for scheduling algorithms
or TDMA-like based
medium access.
The complexity of designing  scheduling algorithms 
   depends on the interference
and propagation models \cite{Goussevskaia2007}. 
Performing graph signal processing tasks   by adopting  link scheduling algorithms
 implies a huge demand on the number of unicast transmissions, which  impacts negatively on the energy available
 at the battery-powered sensor nodes. For this reason, adopting broadcasting
scheduling algorithms to perform graph signal processing tasks is more efficient
since many  links can be activated at the same time.

Broadcasting
scheduling algorithms  
mainly focus on solving the local broadcasting problem.
 In such problem, each node needs to broadcast a message to its neighbors
within  some    local  broadcast  range. 
Local broadcasting was first introduced in  \cite{GoussevskaiaMoscibroda2008}, where the authors propose 
two distributed asynchronous randomized algorithms for the physical SINR model,  with the assumption that
the time is divided in fixed slots.  
The first  algorithm, referred  in this work as LBPIM,
assumes that each of the $N$ nodes knows 
the number of nodes in its proximity  $\Delta_i$ and  can complete a successful broadcast 
with a probability at least $1-1/N^2$ after $O(\Delta_i \log N)$  time-slots. 
In this algorithm, every node 
 decides after about $\Delta_i \log N$ time slots to transmit a packet,  within a certain commun radius,  with a probability $1/\Delta_i$ 
 or remain silent with a probability $1-1/\Delta_i$.
The second  algorithm
has no  knowledge of the number of the nodes in  proximity and  each node can complete a successful local broadcast in $O(\Delta_i  \log^3 N)$  time-slots.
The asynchronous algorithm with  no  knowledge of the number node in proximity was  later   improved in 
 \cite{YuWang2011}, where an algorithm that ensures a successful local broadcast  in $O(\Delta_i   \log^2 N)$  time-slots is proposed.
 In the same work, the authors  also propose two synchronous algorithms that  do not require   the knowledge of $\Delta_i$,
 use a physical carrier sensing and two different transmission powers. They both
   achieve a successful local broadcast
in $O(\Delta_i  \log N)$ time slots.
Later in  \cite{YuHua2012}, the  authors improve the asynchronous algorithm with no knowledge of the number of nodes in proximity
by ensuring a successful local broadcast in $O( \Delta 
\log N + \log^2 N)$ time slots,   where $\Delta$ is the maximum node degree in the network. This algorithm, referred  in this work as  RLBA,
 uses an adjusted clustering-based approach to elect leader nodes, which 
 coordinate the local broadcasting process,
allowing each node to transmit with a constant transmission probability.
In    \cite{Halldorsson2012}, a slightly similar algorithm that achieves a successful local broadcast in 
$O(\Delta  \log N + \log^2 N)$ time slots is proposed.
In addition to that, another algorithm that provides a successful local broadcast in $O(\Delta + \log^2 N)$
time-slots is proposed.  But, it assumes that the nodes can receive acknowledgments from
the neighbors in the broadcast region or use a carrier-sense mechanism 
(measuring the received power from the other nodes even when transmitting
in order to
verify if the signal is above a certain threshold). 
Many   distributed broadcasting algorithms  based on node coloring   and  requiring
a preprocessing stage
have been  also proposed for the SINR model.
  In \cite{YUWang2014}, without the  knowledge
 of the neighborhood, a distributed randomized $\Delta+1$-coloring algorithm with runtime $O(\Delta  \log N + \log^2 N)$  slots
 is proposed.
 This algorithm 
  assumes that nodes
 can adjust their transmission power up to a constant factor.
In \cite{DerbelTalbi2010}, a synchronized distributed $\Delta$-coloring algorithm with runtime $O(\Delta  \log N)$  slots
 is proposed.  This algorithm 
  assumes    that nodes have the  knowledge
of their neighborhood and 
can tune  the transmission power  during the coloring step.
{In \cite{FuchsPrutkin2015},  a synchronized 
distributed $\Delta+1$-coloring algorithm with runtime $O(\Delta  \log N)$  slots
 is proposed. The communication between nodes in this  algorithm that we refer  in our work as SDDC, is based on a similar approach as the one used in LBPIM  \cite{GoussevskaiaMoscibroda2008}}. 
Even though many broadcast distributed algorithms with low time complexity have been  proposed in the literature, they do
not  satisfy all the needs of graph filtering process, which needs to
ensure the accuracy of the graph filtering result.
{In this paper,  we  extend  our previous work in \cite{BenSaadLozano2018}  by  showing that this accuracy can be achieved
by  designing a new cross-layer protocol that controls the PDRs of link connections
in the neighborhood of each transmitter. 
In addition to  that, we  provide a broader analysis of the problem of applying graph filtering in WSNs with asymmetric links,
by considering both forms  of finite impulse response graph filters (node-invariant and node-variant graph filters)
and  showing how to reach an exact unbiased  filtering.
}

 \vspace{-0.2cm}
\section{Background}
Consider  a directed graph  $\mathcal{G}(\mathcal{V}, \mathcal{E})$  with $\mathcal{V}$  a set of $N$ nodes 
and $\mathcal{E}$  a set of   directed edges,
such that if 
there is a link from node  $j$ to node $i$, 
then $(j, i) \in  \mathcal{E}$. 
We define for  any given graph $\mathcal{G}$, the $N{\times}N$ adjacency matrix $\mathbf{A}$, where $a_{ji}=1$ 
if and only if $(j, i) \in  \mathcal{E}$. 
Let the  set of (outgoing)  neighbors of node $j$ be $\Omega_j=\{i \in \mathcal{V} : (j,i) \in \mathcal{E}  \}$.
 We  define the directed Laplacian matrix $\bold L$ of a graph as
$\bold L= \bold D- \bold A$, where
 $\mathbf{D}$  is the diagonal matrix whose non-zero entries are given by the out-degree
$[\mathbf{D}]_{jj}= |\Omega_j|$ \cite{TremblayGoncalvesBorgnat2018}. 
Note that for an undirected graph,
the  Laplacian $\bold L$  
is  symmetric.


On the nodes of $\mathcal{G}$, we can define a graph signal  as a map  $x: \mathcal{V} \to \mathbb{R}$. 
This graph signal  can be denoted by a vector $\mathbf{x}=[x_1,...,x_N]^\top$, whose $i$th entry $x_i$ refers to the signal at node $i$.
 Any  graph $\mathcal{G}$  can be referred  as a so-called
graph shift operator $\mathbf{S}$, which  forms the basis for processing the graph signal and
can be represented as a matrix $\bold S \in \mathbb R^{N  \times N}$. An entry 
 of  $\bold S$ can be  non-zero 
    only if $i=j$ or if $(j, i) \in \mathcal E$.
We can select as shift operator $\bold S$  the
  adjacency matrix $\bold A$, the  Laplacian matrix $\bold L$, as well as their normalized
counterparts or  generalized forms.  

 \vspace{-0.2cm}
 \subsection{Graph filters} 
A graph filter (GF) is  a linear operation $\mathbf{H}$ on an input  graph signal $\mathbf{x}$,  
generating an output  graph signal $\mathbf{y}$. 
We represent a graph filter $\mathbf{H}:{\mathbb{R}^{N}}\rightarrow{\mathbb{R}^{N}}$ 
 by an $N{\times}N$ matrix. 
We can classify the different implementations of GFs into two types: 
Finite Impulse Response (FIR) and Infinite Impulse Response (IIR)   \cite{Sandryhaila2014,SegarraMarques17,ShiFeng2015}.
In this work, we are interested in  one of the central problems in WSNs, 
which is signal denoising,
particularly, we concentrate our focus on the Tikhonov denoising problem using FIR GFs\footnote{Notice that our work can be  extended to other  implementations 
of GFs  performed  over
time-varying networks, where the coefficients do not require  the knowledge of
the network topology, such as in graph signal diffusion \cite{Shuman2013}.
}, 
by taking advantage from the equivalence that exists with  a specific type of IIR GFs, named ARMA \cite{IsufiLoukasSimonetto2017}. 
Next, we introduce the main concepts related to FIR GF and explain its connection to 
ARMA graph filters for Tikhonov denoising. 



\subsubsection{FIR graph filters}

FIR graph filters can be implemented in different ways. 
The two widely used  implementations  are   node-invariant or node-variant\footnote{Recent works have extended these implementations to 
the edge-variant graph filter \cite{CoutinoIsufiLeus2018}.} \cite{SegarraMarques17}:
 
\subsubsection*{Node-invariant graph filters}
 
 Performing  the node-invariant graph filter $\bold H_{inv}$ on the input graph signal $\bold x$ leads to the filter output:
\begin{equation}
 \bold y = \bold H_{\text{inv}} \bold x = \displaystyle\sum_{l=0}^{L} h_{l} \;\bold S^{l} \bold x= \displaystyle\sum_{l=0}^{L} h_{l}  \;\bold x^{(l)}
 \end{equation}
\noindent where $L$ is the filter order, the vector $\bold h=[h_0,..., h_{L}]^\top$ contains the filter coefficients and $\bold x^{(l)}$ = $\bold S^l \;\bold x$ = $\bold S\; \bold x^{(l-1)}$.

%
%
%
%
%
%
%
%
%
%
%
%
%

\subsubsection*{Node-variant graph filters}
 
  Performing the node-variant graph filter $\bold H_{\text{nv}}$ 
 on the input  signal $\bold x$ leads to the output \cite{SegarraMarques17}: 
\begin{equation}
\bold y = \bold H_{\text{nv}} \bold x  = \sum_{l=0}^{L}  \text{diag}(\bold h^{(l)}) \;\bold S^{l} \; \bold x
\end{equation}
\noindent where  the  $N{\times}1$ vector $\bold h^{(l)} = [h_{1}^{(l)},...,h_{N}^{(l)}]^{\top}$ contains  the filter coefficients.

%
%

\subsubsection{IIR ARMA$_{1}$ graph filters}
{ARMA$_{1}$ which denotes an ARMA graph filter  of order one,  is the building block of ARMA graph filter \cite{IsufiLoukasSimonetto2017} and has
as output:} 
\vspace{-0.3cm}
\begin{equation}
 \bold y_{t}= \psi \bold S \; \bold y_{t-1} \;  +  \varphi \; \bold x=( \psi  \bold S )^{t}  \bold y_{0} +\; \varphi  \displaystyle\sum_{\tau =0 }^{t-1} (\psi \bold S)^{\tau}   \; \bold x
\label{eq-ARMA1}
 \end{equation}
where $\varphi$ and $\psi$ are the filter coefficients. 

If $\bold y_{0}=\bold x$,  ARMA$_{1}$  filter provides the same 
 output as that of the node-invariant graph filter  of order $L=T$ with  coefficients $[\varphi, \varphi \psi,..,\varphi \psi^{T-1}, \psi^{T} ]^\top$.
In \cite{IsufiLoukasSimonetto2017},  it is shown that
 ARMA$_{1}$  can recover a signal of interest $\bold v $ from a noisy realization  $\bold x=\bold v+ \bold n$,
 where  $\bold n$ is the noise and with  prior assumption that the graph signal $\bold v $  varies smoothly with  respect to the underlying graph.
 This problem, known as Tikhonov denoising, can be formulated as: 
 \begin{equation}
 \bold v^{*}=  \underset{\bold v \in {\mathbb{R}}^N} {\text{arg min}} \| \bold v -\bold x    \|_2^2  + w \; \bold v^\top  \bold S \bold v 
 \label{eq-Tikhonov-denoising}
 \end{equation}
 
 \noindent  where   
 $w$ is the regularization weight that trades-off smoothness and noise removal and $\bold S= \bold L$.
  The  well-known solution of (\ref{eq-Tikhonov-denoising}) is   $\bold v^{*}=(\bold I + w \bold S)^{-1} \bold x$, which
  can be obtained by   filtering $\bold x$ with  ARMA$_{1}$ and considering $\psi=-w$ and $\varphi=1$ \cite{IsufiLoukasSimonettoLeus17}.




 \vspace{-0.15cm}
\section{Finite inpulse response graph filter analysis}\label{AnalysisSGF}

We consider a WSN modeled as a random graph $\mathcal{G}=(\mathcal{V}, \mathcal{E})$ and serving as a platform to perform graph filtering tasks. 
The WSN is composed by
$N$  sensor nodes randomly and uniformly deployed  over a certain area of interest
and equipped with  an omni-directional antenna.
Depending on the available transmission power, each sensor node has a maximum transmission range $R_B$
up to which it can communicate  with its neighbors. 
Since the WSN can suffer from random topological changes, 
each graph realization  is denoted as $\mathcal{G}_t=(\mathcal{V}, \mathcal{E}_t)$, which represents
the different possible link activations with certain probabilities for each activation.
Let  $\mathcal{G}_0=(\mathcal{V}, \mathcal{E}_0)$ be the particular graph realization
where all the possible
links   are activated simultaneously within the transmission range $R_B$. 
We assume here that  graph filtering is applied over   
time-varying graphs  $\mathcal{G}_t=(\mathcal{V}, \mathcal{E}_t)$, which are  random realizations  at time $t$
of the   graph $\mathcal{G}$, 
where 
the probability of activating a link $(i,j)$ from node $i$ to node $j$ at time $t$ is  $p_{ij}$ ($0<p_{ij}\leq 1$). 
For each graph realization $\mathcal{G}_t=(\mathcal{V}, \mathcal{E}_t)$, it is also assumed that   the set of links $\mathcal{E}_t \subseteq \mathcal{E}_0$  
are activated independently   over the graph and time and generated via an i.i.d. Bernoulli process
with  associated probabilities $p_{ij}$, which will also depend on the protocol that is used at the link layer.
Note that unlike  most of existing works, the  links are allowed  to be   asymmetric {$p_{ij}\neq p_{ji}$}, in order to   consider a realistic  assumption under real conditions in WSNs.
Let  us indicate the connection probability  matrix that reassembles the link activation probabilities $p_{ij}$ by $\bold P \in \mathbb R^{N \times N} $. 
Conducting  graph filtering over   time-varying  random graphs $\mathcal{G}_t$ implies also that
the graph shift operator changes at each time $t$.
Let us indicate  the shift operators associated, respectively, to the  graph $\mathcal{G}_0$, the graph   $\mathcal{G}_t$ at time $t$
and the expected graph $\bar{\mathcal{G}}$, by $\bold S$,  $\bold S_t$ and $\bar{\bold S}$.
We also consider the graph shift operator $\bar{\bold S}$ associated to the expected graph $\bar{\mathcal{G}}$, {given by the entrywise product of the connection probability matrix and the shift operator:}
\vspace{-0.05cm}
\begin{equation}
 \bar{\bold S}=\mathbb{E}[\bold S_t] = \bold P \circ \bold S
\vspace{-0.1cm}
\end{equation}


 By defining the transition matrix of the different graph realisations  
$\bold \Theta(t',t) = \prod_{\tau=t}^{t'}  \bold S_{\tau} \; \text{if} \; t'\geq  t$ 
and $\bold I  \; \text{if} \; t' <  t$ as in \cite{IsufiLoukasSimonetto2017}, the output of a node-invariant   GF  
is given by:
\vspace{-0.15cm}
\begin{equation}
 \bold y_t =\displaystyle\sum_{l=0}^{L} \phi_{l} \;\bold \Theta(t,t-l+1) \; \bold x
\vspace{-0.15cm}
 \end{equation}
\noindent where  $\phi_{l}$ are the filter  coefficients.


 By considering the independence
of  graph realizations, the expected output of the node-invariant  GF  is given by \cite{IsufiLoukasSimonetto2017}:
\begin{equation}
  \bar{\bold y}_{t} = \mathbb{E} \big[ \bold y_t \big]=\mathbb{E} \left[ \displaystyle\sum_{l=0}^{L} \phi_{l} \; \bigg(\prod_{\tau=t}^{t-l+1}  \bold S_{\tau} \bigg) \; \bold x \right]=\displaystyle\sum_{l=0}^{L} \phi_{l}   \; \bar{\bold S}^{l} \; \bold x
\end{equation}

Our main  interest is to get  an unbiased graph filtering by enforcing  $\bold y_t$ to be close on average to
the output $\bold y$. 
Notice that  we introduce a different set of coefficients $\phi_{l}$  instead of  $h_{l}$ in order to reflect our interest
in determining the coefficients $\phi_{l}$ that will achieve  on average the same filter output as if we perform a filter
with coefficients $h_{l}$ 
over the graph $\mathcal{G}_0$. Let consider 
the expected   error (bias), which  can be expressed as:
\begin{equation}
 \bar{\bold e}=\mathbb{E} \big[\bold y_{t} - \bold y \big]=\bar{\bold y}_{t} -\bold y
\end{equation}
\noindent  where if  $\bar{\bold e}=\bold 0$,  unbiased filtering is obtained.

\noindent As shown in the Appendix, 
an unbiased filtering is achieved if the links are established with equal probability $p$  over the random graphs and
the  coefficients  meet the following conditions:
\begin{equation}
\phi_{l}   = p_{ij}^{-l}  \;\;{h_{l}} = p_{ji}^{-l} \;\;{h_{l}}= p_{ii}^{-l} \;\;{h_{l}}={p}^{-l}\; h_{l} \;\; \forall i,j,l.
\end{equation}

\noindent  Next, we  analyze  graph filtering  with node-variant graph filters
 when performed over  random graphs, leading to the output:
 \vspace{-0.1cm}
\begin{equation}
 \bold y_t =\sum_{l=0}^{L} \text{diag}(\boldsymbol \phi^{(l)}) \;\bold \Theta(t,t-l+1) \; \bold x
\vspace{-0.15cm}
 \end{equation}
\noindent where the $N{\times}1$ vector ${\boldsymbol\phi}^{(l)}=[\phi_{1}^{(l)},...,\phi_{N}^{(l)}]^{\top}$ contains the
filter coefficients.

\noindent Consider applying  node-variant graph filter  over time-varying graphs 
with links activated based on $\bold P$. The expected output over the average  graph $\bar{\mathcal{G}}$ for $t \geq L$,
is given by:
\begin{equation}
  \bar{\bold y}_{t} = \mathbb{E} \big[ \bold y_t \big]
                    =\displaystyle\sum_{l=0}^{L} \text{diag}(\boldsymbol \phi^{(l)})  \; \bar{\bold S}^{l} \; \bold x=\displaystyle\sum_{l=0}^{L} \text{diag}(\boldsymbol \phi^{(l)})  \; {(\bold P \circ \bold S)}^{l} \; \bold x
  \end{equation}
\noindent  If the connection probability matrix $\bold P$ has  all entries such that $p_{ij}=p_{ji}=p$,
 we have the following:

\begin{align}
\begin{split}
  \bar{\bold y}_{t} &=\displaystyle\sum_{l=0}^{L}  \text{diag}(\boldsymbol \phi^{(l)})  \; {(\bold P \circ \bold S)}^{l} \; \bold x=\displaystyle\sum_{l=0}^{L}  \text{diag}(\boldsymbol \phi^{(l)})  \; {(( p\; \bold J) \circ \bold S)}^{l} \; \bold x\\
 &=\displaystyle\sum_{l=0}^{L}  \text{diag}(\boldsymbol \phi^{(l)}) \, p^{l} \; {( \bold J \circ \bold S)}^{l} \; \bold x =\displaystyle\sum_{l=0}^{L}  \text{diag}(\boldsymbol \phi^{(l)}) \, p^{l} \; {\bold S}^{l} \; \bold x
\end{split}
 \end{align}\label{node-variant-unbiasedFilter}

\noindent where  $\bold J$ is the $N \times N$ all-ones matrix.

\noindent It can be easily  seen  that if the filter coefficients used over  time-varying graphs
 are chosen such that  $\boldsymbol \phi^{(l)} = p^{-l} \; \boldsymbol h^{(l)}$ and the links are established with the same probability $p$,
this results in an unbiased filtering, as follows:
\begin{align}
\begin{split}
 \bar{\bold e}&=\bar{\bold y}_{t} -\bold y
  =\displaystyle\sum_{l=0}^{L}  \text{diag}(\boldsymbol \phi^{(l)}) \, p^{l} \; {\bold S}^{l} \; \bold x \;- \displaystyle\sum_{l=0}^{L} \text{diag}(\boldsymbol h^{(l)})    \; \bold S^{l} \; \bold x\\
  &=\displaystyle\sum_{l=0}^{L}  \text{diag}(p^{-l} \; \boldsymbol h^{(l)}) \, p^{l} \; {\bold S}^{l} \; \bold x \;- \displaystyle\sum_{l=0}^{L} \text{diag}(\boldsymbol h^{(l)})    \; \bold S^{l} \; \bold x
  =\bold 0\\
\end{split}
\end{align}
\noindent It can be seen that there is no other combination of probabilities and coefficients
that would make bias equal to zero.

\begin{remark}
As it has  been  shown previously, 
equal probabilities for all links allows to reach  an unbiased filtering for both   node-invariant GFs and 
node-variant GFs. 
But, notice that due to interferences and  noise that naturally lead
 to   asymmetric links \cite{ZamalloaKrishnamachari2007,LifengAnish2010}, 
it is impossible to impose equal (or even similar) probabilities for all the links in WSNs.  
  This means that  in practice, we {cannot} obtain exactly unbiased  graph filtering.
  \end{remark}
  
We propose in the next section a solution for  
performing  in practice   graph filtering over random  WSNs.

%


\vspace{-0.1cm}
\section{Graph filtering over random asymmetric WSN}\label{GF_symLinks}

In this section, we present an efficient solution to execute graph
filtering tasks over  random WSNs, so that 
 a bias-variance trade-off   is minimized.

\vspace{-0.25cm}
\subsection{Bias (expected error)}
Since it is impossible in practice to impose equal (or even similar) probabilities for all the links in WSNs to
ensure $\bar{\bold e}=\bold 0$ (exact unbiased filtering), one can determine another way 
to make  the bias as small as possible 
(i.e., $\bar{\bold e} \approx \bold 0$), when performing  graph filtering   with asymmetric links
established based on a connection probability matrix $\bold P$, which has non necessarily equal entries. 
This implies that in order to 
make  the average   graph  filter output over  time-varying graphs
close to the  output of graph  filter applied  over the  deterministic graph $\mathcal{G}_0$, 
one can minimize  
the  Frobenius norm of the filtering   matrix difference, which
 accounts for the  difference between   graph filtering  over the expected graph $\bar{\mathcal{G}}$ and  graph filtering  over the deterministic graph $\mathcal{G}_0$. 

\noindent For node-invariant graph filter, the Frobenius norm of the filtering   matrix difference is given by:
\begin{equation}
 \| \bold B_{\text{inv}} \|_F=\left\| \sum_{l=0}^{L}  \phi_{l}   \; {(\bold P \circ \bold S)}^{l}- \sum_{l=0}^{L}    h_l   \; \bold S^{l}  \right \|_F
\end{equation}

\noindent For node-variant graph filter, the Frobenius norm of the filtering   matrix difference is given by:
\begin{equation}
 \| \bold B_{\text{nv}} \|_F=\left\| \sum_{l=0}^{L} {\text{diag}(\boldsymbol \phi^{(l)}) }   \; {(\bold P \circ \bold S)}^{l}- \sum_{l=0}^{L}   \text{diag}(\boldsymbol h^{(l)})   \; \bold S^{l}  \right \|_F
\label{eq-Dnv}
 \end{equation}

In order to minimize the bias of  graph filtering,  we can find the optimal coefficients that minimize 
these filtering matrix differences by solving the following  optimization
problems, respectively, for node-invariant GF and node-variant GF:  
\begin{equation}
\begin{array}{ll}
\underset{ \{  \phi_{l} \}}{\text{minimize}} \;  \| \bold B_{\text{inv}} \|_F^2   \\
\end{array} \label{eq-minimizing-Dinv}
\end{equation}
\begin{equation}
\begin{array}{ll}
\underset{ \{\boldsymbol \phi^{(l)} \}} {\text{minimize}}   \| \bold B_{\text{nv}} \|_F^2   \\
\end{array} \label{eq-minimizing-Dnv}
\end{equation}

\begin{figure}[t]
\begin{center}
\subfloat[]{\includegraphics[scale=0.315]{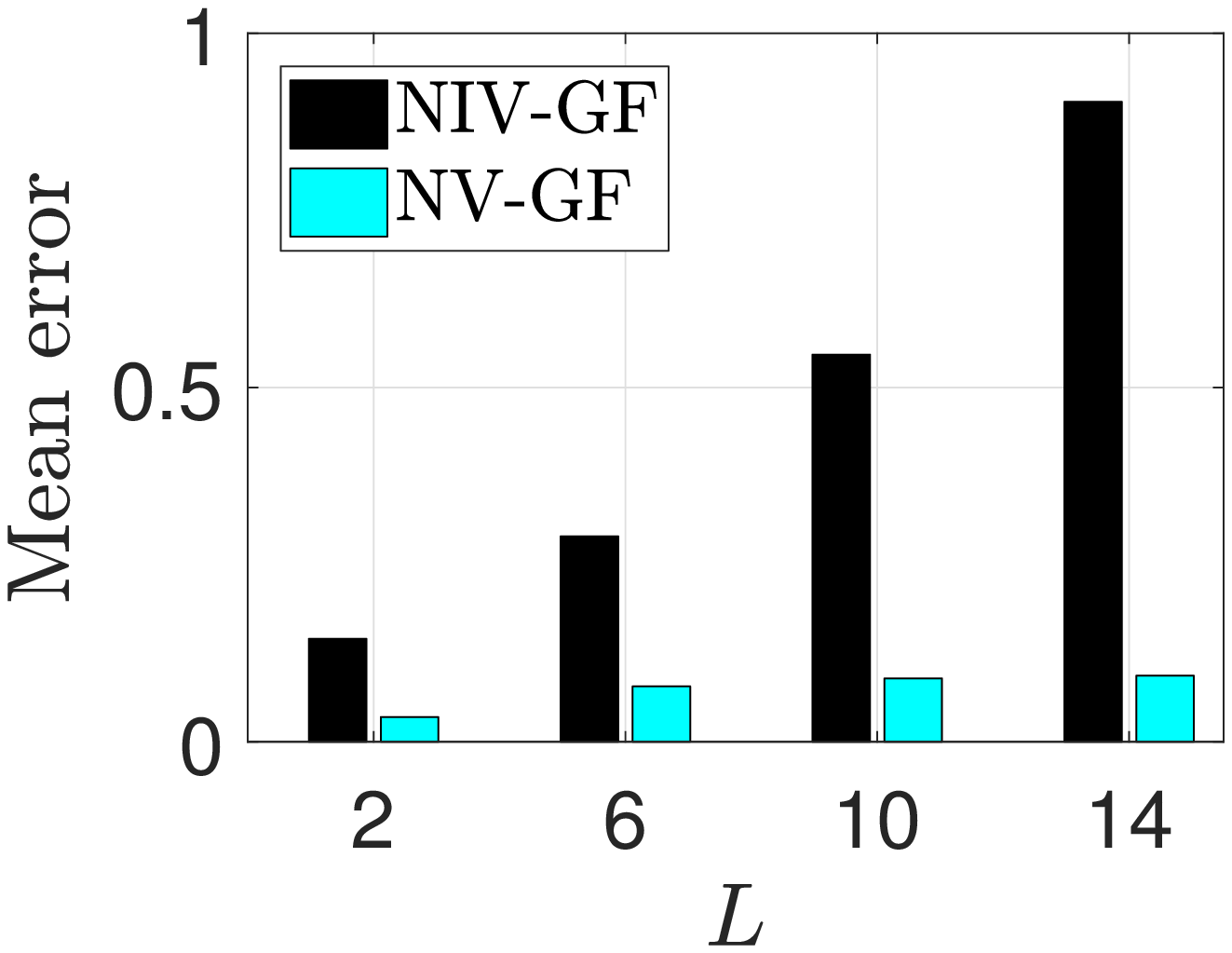}\label{fig-exp-Compare}}
\subfloat[]{\includegraphics[scale=0.315]{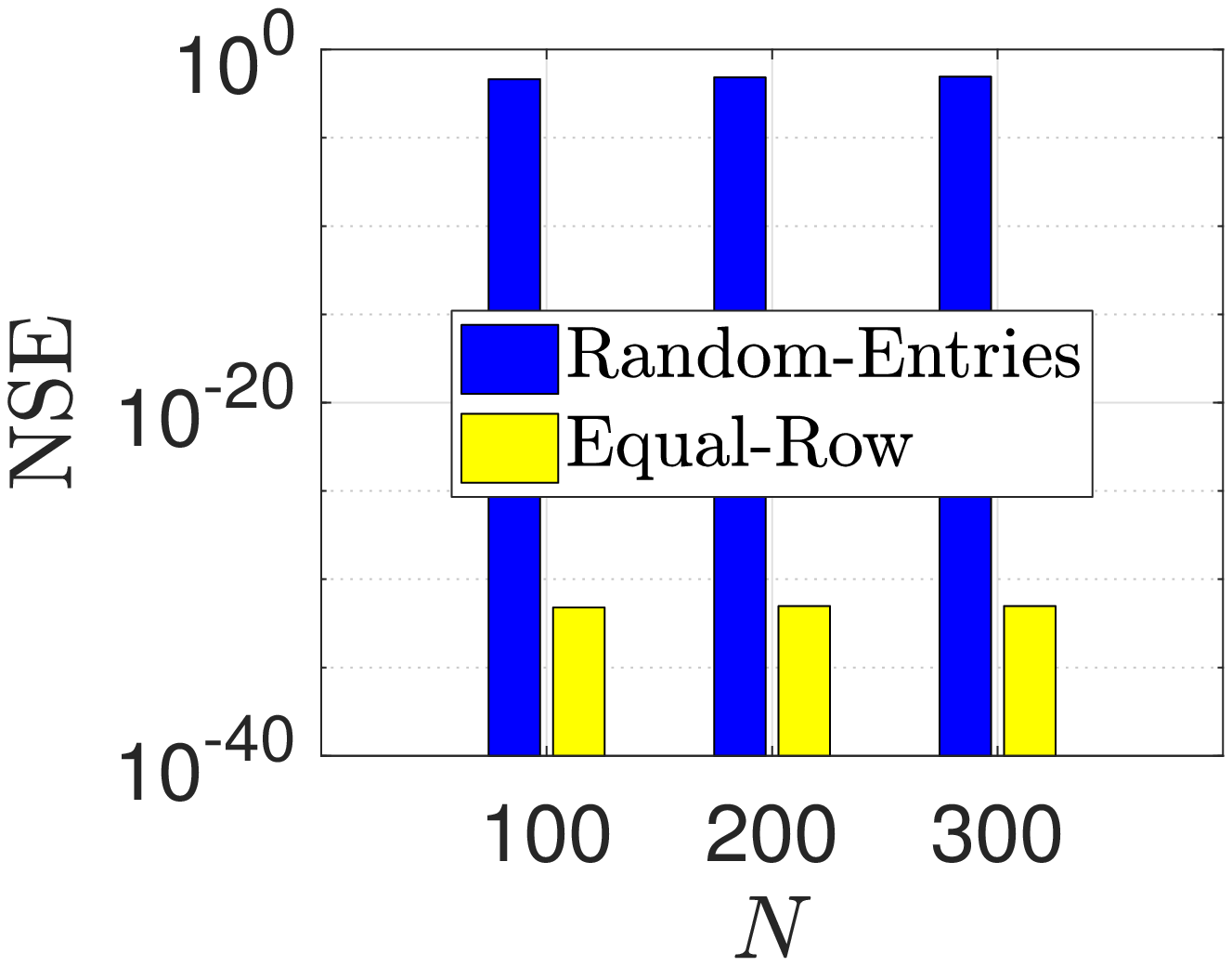}\label{fig-exp-ComparePRow}}
\end{center}
\caption{ (a)  Mean error  among nodes between  performing graph filtering
  over  deterministic graphs and over   time-varying graphs for different values of $L$, with $N=20$,  where
   the probabilities $p_{ij}$   are chosen randomly in $(0,1]$  and the optimal coefficients, obtained  by solving (\ref{eq-minimizing-Dinv}) and (\ref{eq-minimizing-Dnv}), are used.
 (b)~Normalized Squared Error between applying  node-variant graph filters
  over deterministic graphs and  over  time-varying  
 graphs, with random entries and  equalized row entries of $\bold P$, with $L=5$ and $\bold S=\bold A$.
 }
 \label{fig-Compare-ComparePRow}
 \vspace{-0.3cm}
\end{figure}

To decide which type of FIR graph filter is more appropriate  in WSNs under asymmetric links, 
 we compare for both   node-variant and node-invariant GFs, the mean error 
 between  performing graph filtering
  over deterministic graphs and    time-varying  
 graphs, where   the  probabilities of establishing the links are chosen totally randomly and by selecting 
 the  optimal coefficients obtained  by solving (\ref{eq-minimizing-Dinv}) and (\ref{eq-minimizing-Dnv}).
 As it is shown in the example of  Fig. \ref{fig-Compare-ComparePRow}(a),
the node-variant GF  has significantly lower mean error among nodes compared to 
 the node-invariant GF. This is due to the fact that node-variant GFs provide  higher
 degree of freedom for choosing the coefficients. For this reason, in this work,  
  node-variant GFs are chosen to conduct distributed graph filtering tasks in random WSNs.

%
%
%

 Each entry $p_{ij}$ of the connection probability matrix $\bold P$ 
  represents,  in practice, the Packet Delivery Ratio (PDR) of a given link $(i,j)$, 
  which is imposed by  the WSN environment that is affected by the interference and the background noise.
 Since the matrix $\bold P$ may have an impact on the expected error when applying
graph filtering,  in this work, we analyze a method for adjusting
the entries of $\bold P$, without considering the case of enforcing equal probabilities for all the links, since this is
not realistic in practice. A more realistic  approach could be enforcing equal (or similar) entries  for each  row 
 of the matrix, which reflects in practice equal probabilities of link connections
in  the neighborhood of a given transmitter node.

{In Fig. \ref{fig-Compare-ComparePRow}(b),
we analyze  the  Normalized Squared Error 
$\text{NSE}=\|    \bold H_{\text{nv}} {-}   \sum_{l=0}^{L} {\text{diag}(\boldsymbol \phi^{(l)}) }    {(\bold P \circ \bold S)}^{l}  \|^2_F / \left \| \bold H_{\text{nv}}  \right \|^2_F  $
between  applying graph filtering with node-invariant GF
  over deterministic graphs and  over  time-varying  
 graphs, where  the  optimal coefficients are obtained  by solving (\ref{eq-minimizing-Dnv})
 and the  probabilities of establishing the links are adjusted
 so that $\bold P$ has  equal  entries in each row. These adjusted entries   correspond to link connection towards  neighbors that can be reachable within a given  transmission range. 
 The minimum  non zero value in each row 
  is used to equalize these  entries}. 
 As  shown in Fig.~\ref{fig-Compare-ComparePRow}(b), 
 by equalizing  the rows 
 of  $\bold P$, the graph filtering error can be significantly improved.
Therefore, we  assume in this work that every node $i$  uses a probability $q_{i}$ ($0<q_{i}\leq 1$)  to establish a link towards its neighbors, which
is equivalent in practice to having each node use a broadcast communication to transmit a packet to  its neighbors with this probability, and
which should be enforced by an efficient cross-layer distributed MAC protocol. 
In this paper,  the probability  $q_{i}$ represents  the PDR of  a given node $i$ in a WSN. 
From now on, let us define $\bold Q$, which is  the connection matrix $\bold P$ with equalized rows 
 so that
$p_{ij} = q_{i} \; \forall  j \in \Omega_i \; \text{or} \; i=j$.
By establishing the links in the time-varying graph $\mathcal{G}_t$ based on the connection probability matrix  $\bold Q$,
the graph shift operator $\bar{\bold S}$ associated to the expected graph $\bar{\mathcal{G}}$ is given by $\bar{\bold S}=\mathbb{E}[\bold S_t]=\bold Q \circ \bold S$, that is:
\begin{align*}
\begin{split}
\text{If} \; \bold S&= \bold A    \;\text{then}\; \bar{\bold S}=\mathbb{E}[\bold A_t]= \bold Q \circ \bold A\\
\text{If} \; \bold S& =\bold L \;\text{then}\;
\bar{\bold S} =\mathbb{E}[\bold L_t]=\mathbb{E}[\bold D_t - \bold A_t]=\mathbb{E}[\bold D_t] - \mathbb{E}[\bold A_t]\\
&=  \bold Q \circ \bold D -\bold Q \circ \bold A=\bold Q \circ (\bold D-\bold A)=\bold Q \circ \bold L\\
\end{split}
\end{align*}

\noindent This means that by considering the new defined connection probability matrix $\bold Q$ and  in order
to minimize the bias, we need to reduce  the filtering matrix difference given by:
\begin{equation}
  \bold B_{\text{eq}}^{\star} = \sum_{l=0}^{L} \left( {\text{diag}(\boldsymbol \phi^{(l)}) }   \; {(\bold Q \circ \bold S)}^{l}-    \text{diag}(\boldsymbol h^{(l)})   \; \bold S^{l} \right)  
\label{eq_BEq}
 \end{equation}

Next,  we focus  on the variance of the graph filtering, which in addition to the expected error  is essential
 to  control the total Mean Squared Error (MSE).

\vspace{-0.45cm}
\subsection{Variance}
We focus our analysis on 
the average variance across the nodes, which can be expressed as:
\begin{equation}
\overline{var}  [\bold y_t]=\big(\text{tr}\big(\mathbb{E}[\bold y_t {\bold y_t}^{H}]-\mathbb{E}[\bold y_t]\mathbb{E}[\bold {y_t}]^{H}\big)/N
\label{eq_variance}
\end{equation}
The main result related to the average variance is given by the following proposition.
\begin{proposition}
Consider a node-variant graph filter  operating over time-varying networks  with  links activated based on a connection probability matrix $\bold Q$. 
The average variance across the nodes of the graph filter is upper bounded by:
\vspace{-0.1cm}
\begin{align}
\begin{split}
\overline{var}  [\bold y_t]  & \leq \frac{{\|\bold x \|}^2}{N}   \; \bigg (\displaystyle\sum_{l=0}^{L}   \vartheta_l  \bigg )^2  
 \end{split}\label{bound_variance}
\end{align}
\vspace{-0.05cm}
\noindent where   
$\vartheta_l=\rho^{l}   \; \| {{\text{diag}(\boldsymbol \phi^{(l)})}} \|_2$ and 
$\rho$ is an upper bound of the spectral norm of $\bold S$. 
\end{proposition}

 \textit{Proof}: See Appendix. 

\vspace{-0.2cm}
\subsection{Minimizing the bias-variance trade-off}\label{minimizing-MSE}
In order to ensure a total control on the overall MSE,
we propose in this section to find  the optimal coefficients that  minimize a bias-variance trade-off 
  through solving the  convex optimization problem:
\vspace{-0.25cm}
\begin{equation}
\begin{array}{ll}
\underset{ \{\boldsymbol \phi^{(l)} \}} {\text{minimize}}   \left  \| \bold B_{\text{eq}}^{\star} \right \|^2_F   + \mu \;\; \bigg (\displaystyle\sum_{l=0}^{L}   \vartheta_l  \bigg )^2   \\
\end{array} \label{eq-minimizing-MSE}
\end{equation}
\noindent where $\mu$ is a weighting factor trading-off  the bias and the upper-bound variance.  
Note that due to the fact that the 
positive term ${\|\bold x \|}^2/N$  does not have an influence on the choice of the coefficients
and our  main focus is  to find the coefficients without knowing  the input  signal, this term is omitted. 


Notice that the optimal coefficients that  minimize  the bias-variance trade-off, when performing distributed graph filtering tasks
over random WSNs, depend on the PDRs of the nodes through the connection probability matrix $\bold Q$.
We will
focus  in   Section \ref{section_DCSA} on how these PDRs can be determined and enforced
at the MAC layer by designing
 a  cross-layer protocol.

\vspace{-0.1cm}
 \section{Cross-layer distributed scheduling algorithm for Graph Filtering}\label{section_DCSA}
  In this section, we analyze how to enforce  
  the  PDRs at the nodes in a WSN,   
  in order to ensure an accurate graph filtering, through the control of  the bias and the variance. 
 As we show in this section, these  PDRs  can be enforced at the MAC layer by designing  a cross-layer distributed scheduling algorithm to be adopted
during the communication  exchanges among the nodes  performing the graph  filtering tasks\footnote{Note that this work can also be easily extended to  graph
filtering tasks that account for  time-varying input signals \cite{BenSaadLozano2019spawc}.}.

\noindent We assume that the  time is divided into 
slots  of a fixed duration $\tau$. At each time slot, a certain number of nodes
are allowed to communicate. Our goal is to design a specific scheduling algorithm where the communicating nodes execute the graph
filtering steps, resulting into a  high  accuracy of the filtering process.
Without loss of generality\footnote{Note that this work does not  depend on the specific physical layer of IEEE 802.15.4  and can be easily extended
to other physical layers by using other methods to compute the BER.}, we assume  that the WSN uses a similar  physical layer as the one corresponding to the standard IEEE 802.15.4 with
the 2.4 GHz ISM band\footnote{2.4 GHz ISM  band  has the advantage of being used worldwide without any
limitations on applications.}.
The Bit Error Rate  (BER),
when  node $i$ is transmitting to node $j$  over the link $(i,j)$, 
 is given by \cite{IEEE202.15.4}:  
\vspace{-0.2cm}
 \begin{equation}
  \text{BER}_{i,j}= \displaystyle \varsigma_1 \sum^{\varsigma_2}_{k=2} (-1)^k  e^{\varsigma_3 \; \text{SINR}_{ij} \; \big(\frac{1}{k}-1\big)}
\label{eq-ber}
 \end{equation}
 { where   ${\text{SINR}}_{i,j}$ is  the received signal to interference plus noise ratio at node $j$
 and 
 $\varsigma_1$, $\varsigma_2$ and $\varsigma_3$ are  constants}\footnote{{Note that the constants are  based on a realistic analytical model from IEEE 802.15.4 \cite{IEEE202.15.4}}.} { equal, respectively, to $\frac{1}{30}$, 16 and 20 as stated in \cite{IEEE202.15.4}}.

The PDR of transmitting a  packet of length $z$-bits over a link $(i,j)$
 is given by:
\begin{equation}
 \text{PDR}_{i,j}=  (1 -  \text{BER}_{i,j})^z
 \label{eq-pdr}
\end{equation}
\noindent  assuming that the bit errors occur independently across the $z$-bits of the packet.


Under a physical interference model, the SINR of any link depends on the received signal strength level, which
is related to the transmitting power level, the  distance between the receiver and the transmitter,
and the signal propagation environment. 
In this paper, in order to estimate the  SINR, we adopt 
the log-distance  path-loss propagation model as in \cite{GoussevskaiaHalldorsson2014}.
Including fading to this  model does not significantly affect the performance in real scenarios,
as shown in \cite{AlonsoCeladaAsensio2013}.
According to this model, the received power ${P}_{i,j}$ at a node $j$ from a transmitter node
$i$ over a  link $(i, j)$ can be expressed as:
\begin{equation}
 {P}_{i,j}=\frac{{P}}{ d_{i,j}^{\nu}}
\end{equation}

\noindent where  $d_{i,j}$ is the distance between both nodes $i$ and $j$, 
 $\nu$ is the path
loss exponent, and ${P}$ is the  transmitter power,  which is assumed to be the same for all nodes. 

The ${\text{SINR}}_{i,j}$  of a link $(i, j)$ 
is equal to 
the received power  at node $j$ from node $i$ divided by the sum of received
powers at node $j$ from all other concurrent transmissions,  plus noise.
\begin{equation}
 \text{SINR}_{ij}=\frac{ \; {P}_{i,j}} { I_j +\mathscr{N}_0} 
\end{equation}

\noindent where $\mathscr{N}_0$ is the background noise,  which we assume to be constant and known\footnote{In practice, this can be estimated through calibration.}, 
 and $I_j$ is the interference experienced at node  $j$, given by
$I_j=\sum_{u \in \mathcal{V},\; u \neq i }^{}    \;{P}_{u,j}$.

In the \text{SINR}-based physical model \cite{GuptaKumar2000}, 
the successful reception
of a packet sent by a node $i$ to a node $j$ is achieved if
the SINR at $j$ is higher than a certain value of \text{SINR} threshold   $\kappa$, 
which can be chosen  to guarantee  a small BER: 
  \begin{equation}
   \text{SINR}_{ij}=\frac{ \; \displaystyle \frac{{P}}{ d_{i,j}^{\nu}}} {  \sum_{u \in \mathcal{V},\; u \neq i }^{}    \; \; \frac{{P}}{ d_{u,j}^{\nu}}  +\mathscr{N}_0}      \geq \kappa 
   \label{ineq-threshold}
  \end{equation}
  
\noindent According to the physical interference model, a packet can be correctly received
even if there are a single or multiple simultaneous transmitter
nodes  in the neighborhood of a receiver node $j$, as far as  inequality   (\ref{ineq-threshold}) holds.
  

{Next,  we introduce  some definitions of specific areas  that will be used in the design of our cross-layer
distributed scheduling algorithm. The frequently used notations and terminologies are summarized in  Table \ref{table_notations}}. 
 
\begin{definition}
The maximum transmission radius
$R_\text{m}$  is defined as the maximum distance up to which a
packet sent by a transmitter can be  received  by every node inside the associated circular area of
radius  $R_\text{m}$,
in absence of interference, which  is given by:
 \begin{equation}
  R_\text{m} =\left( \frac{{P}}{\kappa \;\mathscr{N}_0 }  \right)^{ \frac{1}{\nu}}.
 \end{equation}
 \end{definition}
 
\begin{definition}
The  broadcast range $R_\text{B}$ of every
 node with transmitter power $P$,  is the distance up to which the node  intends to broadcast
its messages, when performing graph filtering steps. 
In general, $R_\text{B}$ is lower than  $R_\text{m}$ due to the presence of interference and can be  expressed as:
 \begin{equation}
  R_\text{B} = \chi \; R_\text{m}
 \end{equation}
 
 \noindent where $0<  \chi  < 1$ and the set of nodes inside the broadcast region  of range $R_\text{B}$ when a given node $i$ is the transmitter, is denoted as $\Delta^B_i$. The neighbors of node $i$
are the nodes within the range $R_\text{B}$.
\end{definition}
\noindent Given a certain deployment of the  nodes, 
the value of  $\chi$  must be selected so that it 
ensures  a connected network, that is,  there is a  path between every pair of nodes. 
We will show in Section \ref{algo-CDSA} how this requirement can be achieved.

  \begin{definition}
A successful  broadcast for a node $i$ is defined as  a transmission
of a message during the graph filtering process, such that it is successfully received by
all receivers $j$ located in the  broadcast region within  range $R_\text{B}$, where the condition for a successful reception is given by
(\ref{ineq-threshold}).
\end{definition}

  \begin{figure}[t]
\begin{center}
\scalebox{0.34}{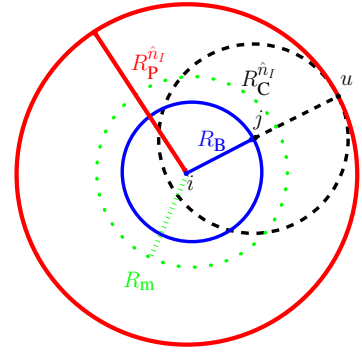}
\end{center}
\caption{Collision area and preventing area with  an interfering node $u$  ($\hat{n}_I=1$). Notice that when node $i$
is transmitting all nodes inside the blue region are receiving and $j$ and $u$ can be any point in the perimeters (blue and red).}
\vspace{-0.25cm}
\label{fig-rangepreventing}
\end{figure}

For any transmitter node $i$ in the network, let us consider  
  the worst case of interference   experienced by the farthest   receiver $j$ located at any point on the 
perimeter of the broadcast region, i.e., $d_{ij}= R_\text{B}$, and let us assume that there are   
$\hat{n}_I$  interfering  nodes  whose distances to 
the receiver are   lower bounded by a certain distance $R_\text{C}^{\hat{n}_I}$ i.e., $d_{uj}\geq   R_\text{C}^{\hat{n}_I}, \;\forall \; u \in \mathcal{V},\; u \neq i,\; u \neq j$,
as illustrated in Fig.~\ref{fig-rangepreventing}. 
From   (\ref{ineq-threshold}), we can write that the worst case  $\text{SINR}^{(\text{wst})}$ is given by:
      \begin{equation}
 \text{SINR}^{\text{wst}} =\frac{ \displaystyle \frac{{P}}{ R_\text{B}^{\nu}}} {  \hat{n}_I \frac{{P}} {   \big( R_\text{C}^{\hat{n}_I}\big)^{\nu} }   + \mathscr{N}_0}    \geq \kappa \;    
   \label{SINR-WST-eq}
   \end{equation}
\noindent and where, as we  explain later in Section \ref{algo-CDSA}, $\hat{n}_I$  is  actually an estimation of 
the number ${n}_I$ of interfering  nodes,
which is computed based on $\hat{N}$ an estimation of the  total number of nodes in the network.
    
\begin{definition}
Given a transmitter node with a certain broadcast region of radius $R_\text{B}$, and any receiver node placed at  the 
perimeter of the broadcast region, the collision area is 
the circular  area  centered at that receiver, with a radius  $ R_\text{C}^{\hat{n}_I}$ given by:
 \begin{equation}
   R_\text{C}^{\hat{n}_I}= \bigg( \frac{\hat{n}_I \;\kappa \;{P}\;\;  R_\text{B}^{\nu} }{ {P}   - \kappa  R_\text{B}^{\nu} \;\mathscr{N}_o   }   \bigg )^{\frac{1}{\nu}}.
\label{equation_Rc}
  \end{equation}
  \end{definition}
  \noindent Note that $R_\text{C}^{\hat{n}_I}$ represents the smallest distance (from any intended receiver $j$) at which we can have other $\hat{n}_I$ transmitters
  interfering, while still having successful communications at the receivers inside the broadcast region of  radius $R_\text{B}$  (see Fig. \ref{fig-rangepreventing}). 
  Notice that this radius grows with $\hat{n}_I$,
  implying that more protection to the receivers is imposed when there are more interfering transmitters.

\begin{definition}
The preventing area   for a  transmitter node is defined as the circular region
centered at that node with the radius  $R_{P}^{\hat{n}_I}$  given by:
 \begin{equation}
  R_{P}^{\hat{n}_I} =  R_\text{B} +  R_\text{C}^{\hat{n}_I}
  \label{equation_Rp}
 \end{equation}
 \end{definition}
 \noindent The preventing area is illustrated  in Fig. \ref{fig-rangepreventing}.
 The set of nodes inside the area of radius $R_{P}^{\hat{n}_I}$ for a  transmitter node $i$ is denoted by $\Delta^P_i$.
 The ring formed by the outer radius $R_{P}^{\hat{n}_I}$ and inner radius $ R_\text{B}$ contains
  the locations of the nodes that are
 responsible
for the most significant part of interference experienced
by the  neighbors of node $i$, located inside the broadcast area of radius $ R_\text{B}$, when receiving from  a transmitter node $i$.

\begin{algorithm}[h]
\begin{algorithmic}[1]
\small
 \REQUIRE $\hat{N}$,   $\text{ID}_i$, $pos_i$, $\tau=0$, $n_{{t}_x}=0$, $t_{\text{out}}$, $\text{timer}=0$, $pkt_{\text{sent}}$ 
\hspace{0.20cm}
\Comment{\tt  /*$pkt_{\text{sent}}$ controls the number of packets sent by feasible nodes*/}
\STATE $\tau=\tau+1$; $\mathcal{F}[\tau]=\emptyset$; $pkt_{\text{sent}}=0$ 

 
 \STATE $state=\text{isActivateNode(}\text{ID}_i \text{)}$
 \IF { $state$ == \text{'active'}}
 \STATE  $\hat{n}_I= \hat{N} - n_{{t}_x} -1 $
 \STATE $ \text{send packet with } pos_i \text{ and } \hat{n}_I$
  \STATE $ \text{reset and trigger}$ $\text{timer}$
  \WHILE {$ \text{timer} < t_{\text{out}}$    }
  \IF {$\text{receiving  packet from feasible nodes}$}
  \STATE $\text{update } \mathcal{F}[\tau] $
 \STATE  $ \text{check and remove conflicting feasible nodes in } \mathcal{F}[\tau]$
  \ENDIF
   \ENDWHILE
 \IF {$| \mathcal{F}[\tau]| ==\hat{n}_I$ and $\hat{n}_I > 0$}
 \STATE $\mathcal{T}_{t_x}^{(\tau)}=\mathcal{F}[\tau]    \cup \{i\}$; add  positions of $\mathcal{T}_{t_x}^{(\tau)}$ in ${\xi}^{(\tau)}$
 \STATE $n_{{t}_x}=n_{{t}_x}+| \mathcal{F}[\tau]|+1$
  \STATE $ \text{send allocated nodes } \mathcal{T}_{t_x}^{(\tau)}$,  ${\xi}^{(\tau)}$, and  $n_{{t}_x}$; $\text{exit( )}$

 \ELSE

   \IF { $| \mathcal{F}[\tau]| \neq \hat{n}_I$ and $\hat{n}_I > 0$}

 \STATE   $\hat{n}_I=\hat{n}_I-1$
  \STATE  $\text{goto line 5}$ 
 \ELSE  
 \STATE $\mathcal{T}_{t_x}^{(\tau)}= \{i\}$; ${\xi}^{(\tau)}= \{pos_i\}$
  \STATE  $n_{{t}_x}=n_{{t}_x}+1$
 

   \STATE $ \text{send allocated nodes } \mathcal{T}_{t_x}^{(\tau)}$, ${\xi}^{(\tau)}$ and  $n_{{t}_x}$; $\text{ exit( )}$
   \ENDIF
   
 \ENDIF

\ELSE 
 
 \IF {$\text{receiving } pos_i\text{, } \hat{n}_I$ \text{from active node and} $pkt_{\text{sent}}==0$}
\STATE $\text{compute }  R_\text{C}^{\hat{n}_I} \text{ using (\ref{equation_Rc})} $
  \STATE           $R_\text{P}^{\hat{n}_I} = R_\text{B}+ R_\text{C}^{\hat{n}_I} $ 
\STATE   $\text{reset and trigger}$ $\text{timer}$

  \WHILE { $\text{timer} < t_{\text{out}}$}
  \IF {$\text{overhearing  packet from feasible node}$}
\STATE $\text{store information of overhearing feasible nodes}$
\ENDIF
\ENDWHILE
  \IF { $\text{\textit{Prevent-Condition} satisfied and no conflicts}$}
\STATE $\text{send packet(}\text{ID}_i, pos_i, \textit{'feasible'} \text{) to active node }$
\STATE $pkt_{\text{sent}}=1$
\ENDIF
\ENDIF
 \IF {$\text{receiving } \mathcal{T}_{t_x}^{(\tau)}$, ${\xi}^{(\tau)}$ $\text{and}$  $n_{{t}_x}$}
\IF {$i \in \mathcal{T}_{t_x}^{(\tau)}$}
\STATE  $                                                                                                                                                                                                                                                                                                                                                                                                                                                                                                                                                                                                                                                                                                                                                                                                                                                                                                                                                                                                                                                                                                                                                                                                                                                                                                                                                                                                                                                                                                                                                                                                                                                                                                                                                                                                       \text{exit( )} $ 
\ELSE 
\STATE    $\text{goto line 1}$
\ENDIF
\ELSE 
\STATE $\text{goto line 28}$
\ENDIF

\ENDIF

\normalsize
\end{algorithmic}
\caption{\small{CDSA algorithm running at each node $i$}}\label{algorithm}
\end{algorithm}

\begin{figure}[t]
\begin{center}
\subfloat[Slot $\tau$, $\hat{n}_I$=3]{\hspace{-0.44cm}\includegraphics[scale=0.235]{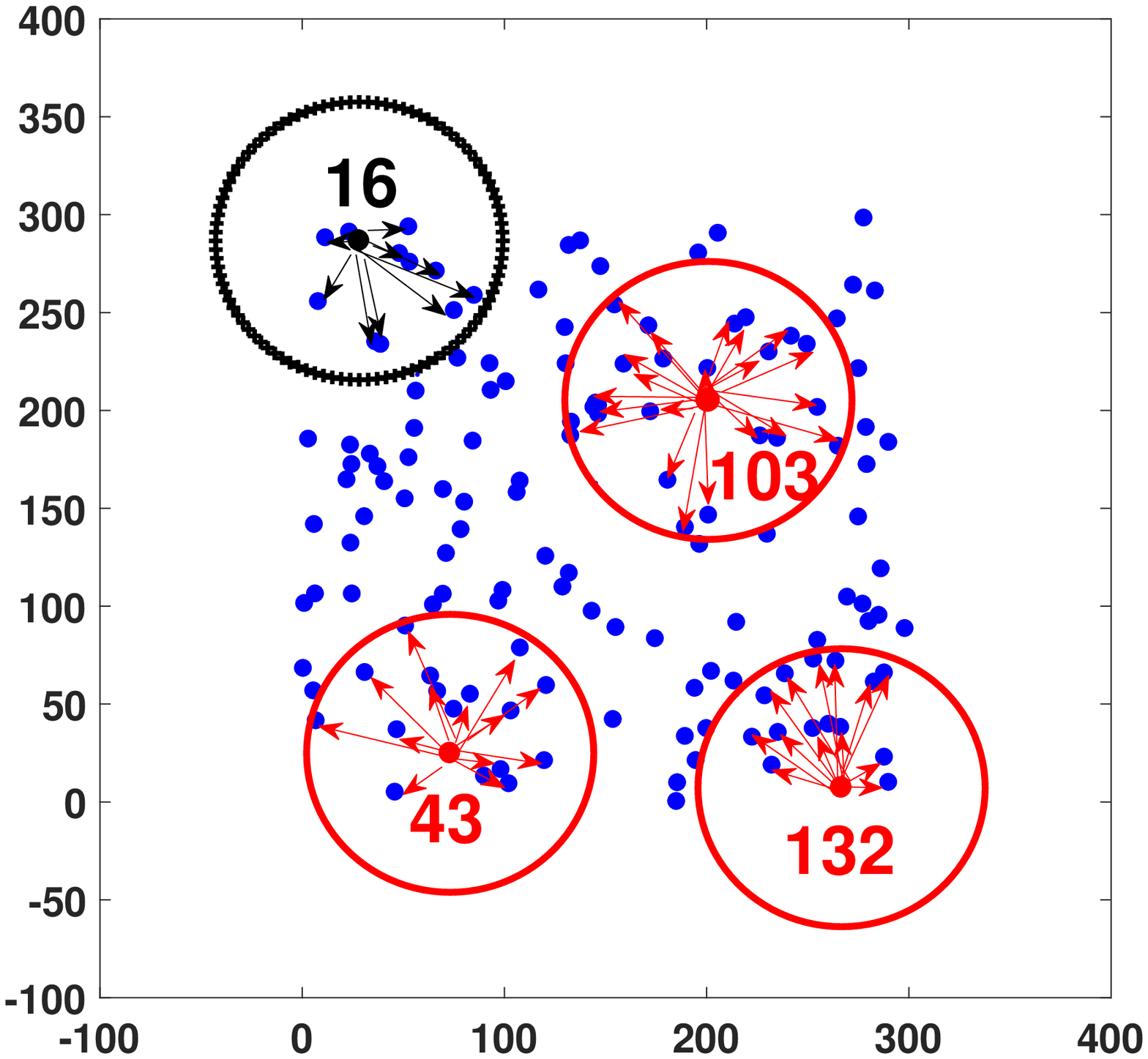}}\hspace{-0.44cm}
\subfloat[Slot $\tau$+1, $\hat{n}_I$=2]{\includegraphics[scale=0.235]{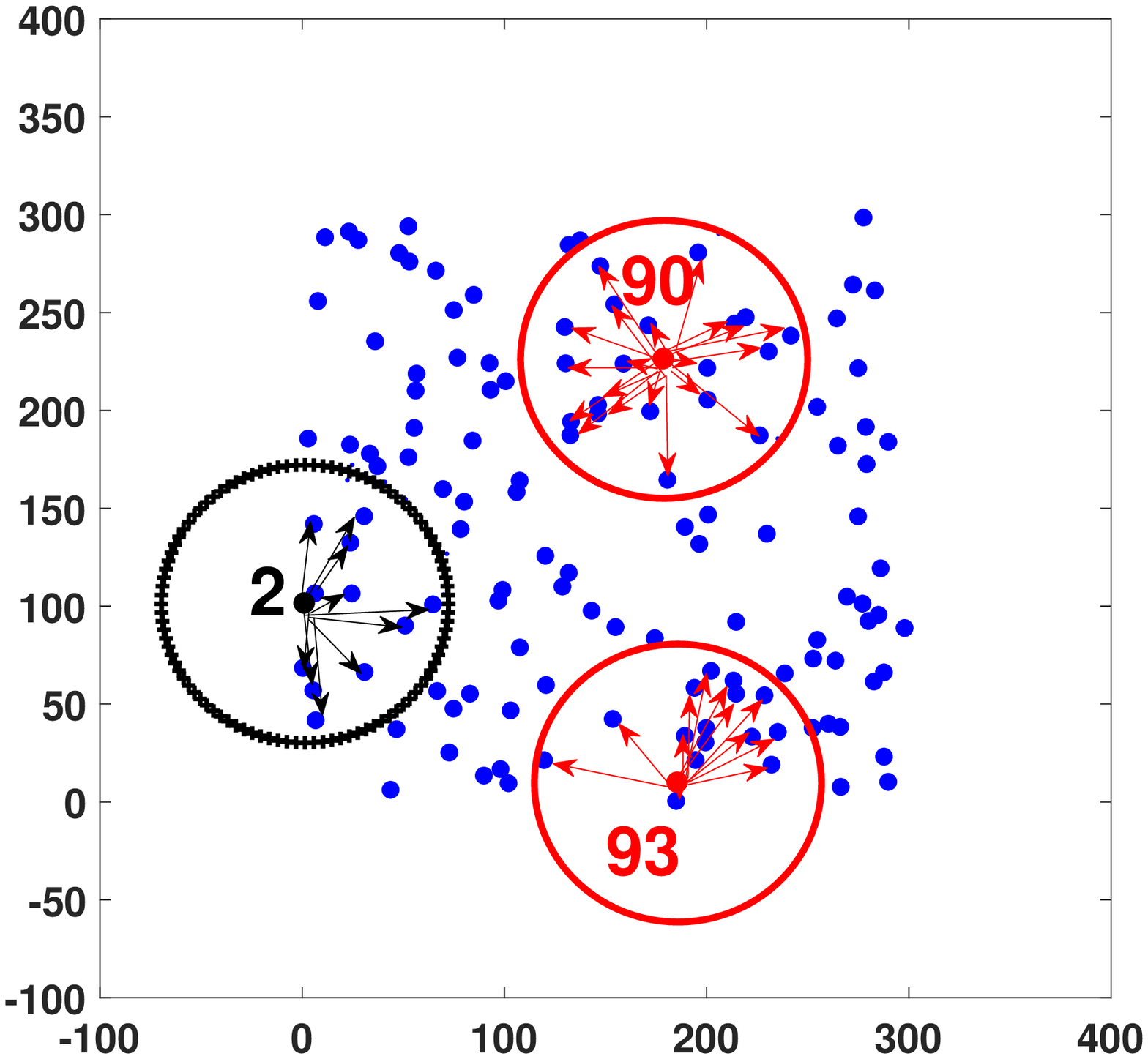}}
\end{center}
\caption{Example of slot allocation  in our CDSA algorithm, where $\mathcal{T}_{t_x}^{(\tau)}=\{16, {43}, {103}, 132\}$ and $\mathcal{T}_{t_x}^{(\tau+1)}=\{{2},{90},{93}\}$. 
The nodes are represented by dots. The  preventing area of the active node is shown with   a  dashed black circle whereas the preventing regions
corresponding to the feasible nodes are indicated 
by plain red circles.}
\label{fig-allcation-slots}
 \vspace{-0.2cm}
\end{figure}

\begin{small}
\begin{table*}[!t]
\caption{{Frequently used notations}}
\label{table_notations}
\centering
{
\begin{tabular}{c|c||c|c||c|c}
\hline
\bfseries Parameter & \bfseries Description &\bfseries Parameter & \bfseries Description&\bfseries Parameter & \bfseries Description\\
\hline
$L$       & Filter order         & $\bold S$   & Shift operator & $\bold x$ &  Input graph signal\\
$R_\text{m}$       & Max. transmission range          & $N$   & Nr. of nodes & $\bold Q$ & Connection probability matrix \\
$R_\text{B}$        & Broadcast range            & $P$   & Transmission power&  $\tau$  & Time slot\\
$\hat{n}_I$         &  Nr. of interfering nodes   & $\nu$ & Path loss exponent& $T_s$& Nr. of allocated slots\\
$R_{P}^{\hat{n}_I}$ &  Range of preventing area   & $\kappa$  & SINR threshold&  $\mathcal{T}_{t_x}^{(\tau)}$& Set of transmitters at slot $\tau$\\
 $R_{C}^{\hat{n}_I}$&  Range of collision area   & $\mathscr{N}_0$& Background noise & $n_{{t}_x}$& Nr. of current allocated slots\\
\hline
\end{tabular}
}
\end{table*}
\end{small}

\vspace{-0.3cm}
 \subsection{Cross-layer Distributed Scheduling  Algorithm (CDSA)}\label{algo-CDSA}
 Given a set   of   $N$  transmitter nodes  that  intend to broadcast a packet inside an area of radius $R_\text{B}$, in order to perform the graph filtering steps by
means of communication exchanges,  our  goal  
is to design a scheduling protocol 
that activates simultaneously, at each time slot $\tau$, a disjoint subset of transmitters $\mathcal{T}_{t_x}^{(\tau)}   \subset \mathcal{V}$ 
such that $\cup^{\tau=T_{s}}_{\tau=1} \mathcal{T}_{t_x}^{(\tau)} = \mathcal{V}$, and  by ensuring at each slot $\tau$
 that the SINR of all receivers inside each broadcast region of each activated transmitter 
 is higher that $\kappa$, and where the $T_{s}$ is the total number of slots.
 Therefore, the aim is to schedule all these requests 
  in a small number of slots $T_{s}$ and satisfying:
   \begin{equation}
 \text{SINR}^{(\tau)}_j=\frac{ \frac{{P}}{d_{i,j}^{\nu}}} {  \displaystyle \sum_{u \in  \mathcal{T}_{t_x}^{(\tau)} \setminus \{i\}} \frac{{P}} {   {d}^{\nu}_{u,j} }   + \mathscr{N}_0}    \geq \kappa, \forall j \in \Delta^B_i,   \forall i \in  \mathcal{T}_{t_x}^{(\tau)}, \forall \tau
  \label{condition-SINR}
   \end{equation}
   \noindent so that all the graph filtering steps are carried out as quickly as possible, maximizing time efficiency.

On the one hand, we are interested in ensuring successful simultaneous broadcasting  of the transmitters scheduled at each of the time slots and on the other hand, in order to ensure accurate filtering, our protocol should
also control, at each transmitter, the PDRs of its corresponding neighbor nodes.

We assume that each node $i$ has an unique identifier ${\text{ID}_i}$ and  knows its position  $pos_i$ by using a geo-localization system or 
 acquiring its location during the initial network  setup,  for instance  as  described in \cite{KuriakoseSandeep2014}.
  Every node can also estimate its distance to  its neighbors  inside the broadcast area
by means of  exchanging  information locations or using  signal detection  techniques, such as in \cite{ZhaoXiHe2013}. Moreover,
 every node can  determine an estimate 
 $\hat{N}$ of the total number  of   nodes in the network by using   one of the well known  distributed  algorithm for counting the number nodes in WSNs, based on consensus 
 (e.g., \cite{ZhangTepedelenliox011Flu2017}). 
As explained next, the accuracy of the estimated total number of nodes  does not have a significant impact on the allocation of slots obtained by our proposed scheduling algorithm since this value is only used as a starting point in the execution of the algorithm.  

In our proposed cross-layer scheduling algorithm, which we call CDSA, 
during  the  allocation of a certain time slot $\tau$,
the nodes are classified into one of these three sets, each one corresponding  to a certain state:

\begin{itemize}
 \item \textit{Active} node  \big($\mathcal{A}[\tau]$\big): When a single node is activated to initiate the allocation of transmitter nodes for the slot $\tau$, it belongs to this state.
 \item  \textit{Feasible} nodes \big($\mathcal{F}[\tau]$\big): When a  node has 
 its  preventing area of radius $R_P^{\hat{n}_I}$ outside
the preventing area of the active node, it belongs to this state. This  implies  that this node can  potentially be selected as a transmitter in  the slot $\tau$. 
 \item  \textit{Candidate} nodes \big($\mathcal{C}[\tau]$\big): Every  node that is neither in \textit{Active} or \textit{Feasible} state and that has not yet been  allocated a slot, belongs to this state.
\end{itemize}

 \begin{figure*}[t]
\begin{center}
\subfloat[Mean error  with   increasing values of $q_{i}$ for different GFs]{\includegraphics[scale=0.32]{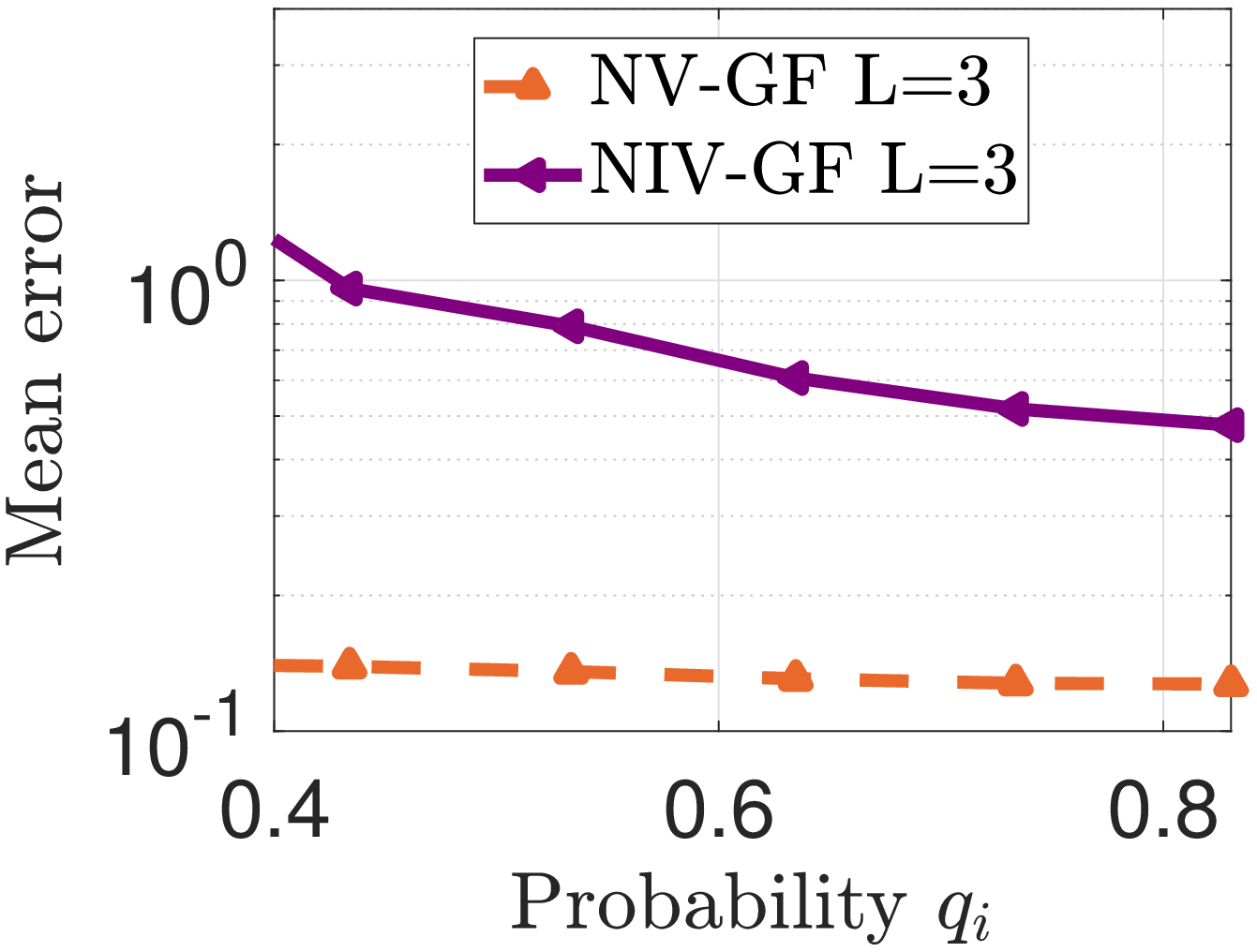}}\hspace{0.02cm}
\subfloat[${{\overline{\sigma}}_{\bold e}^{2}}$ with   increasing values of $q_{i}$ for  different  GFs]{\includegraphics[scale=0.32]{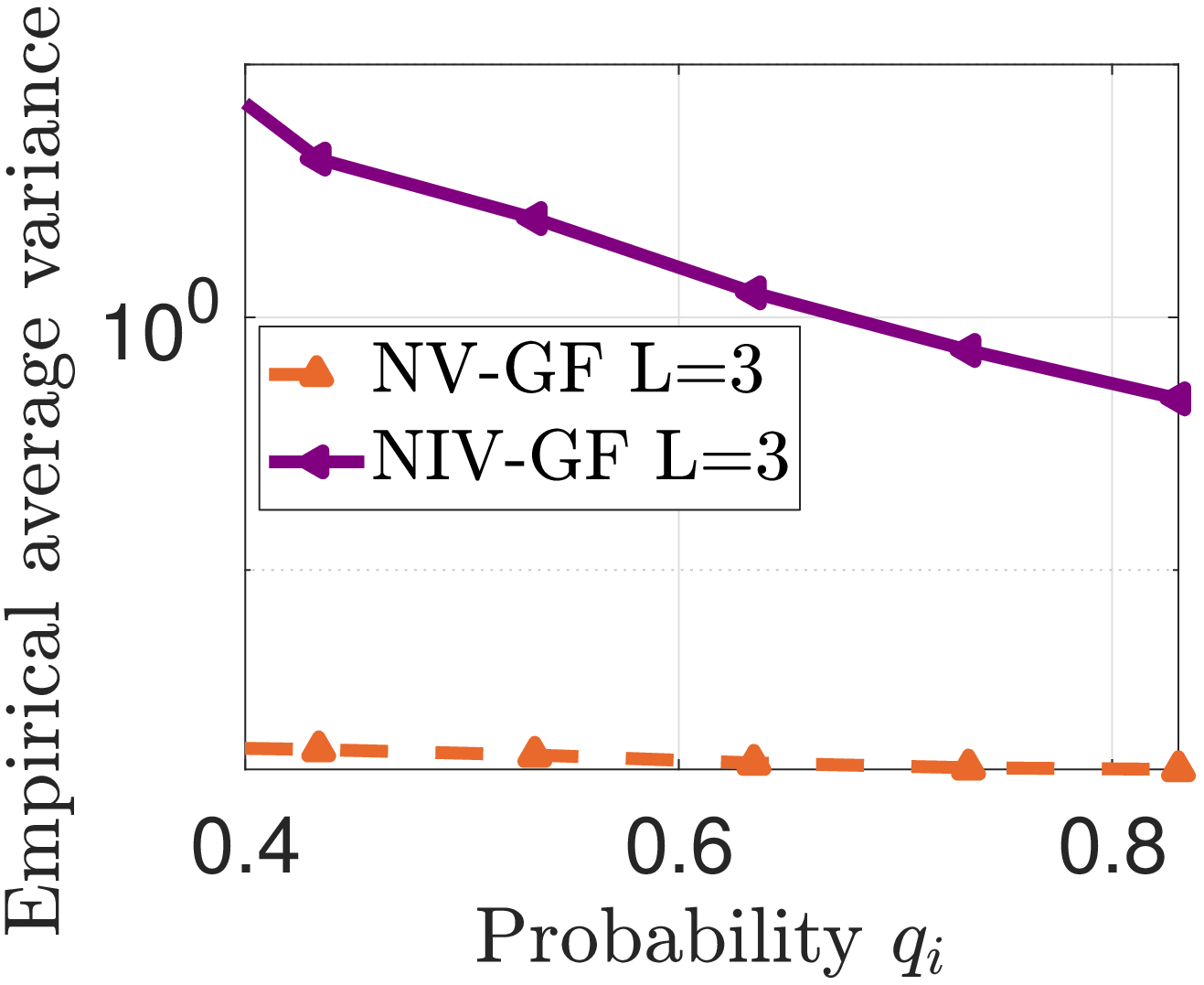}}\hspace{0.02cm}
\subfloat[Mean error  with   increasing values of $q_{i}$ for  different  $L$  ]{\includegraphics[scale=0.32]{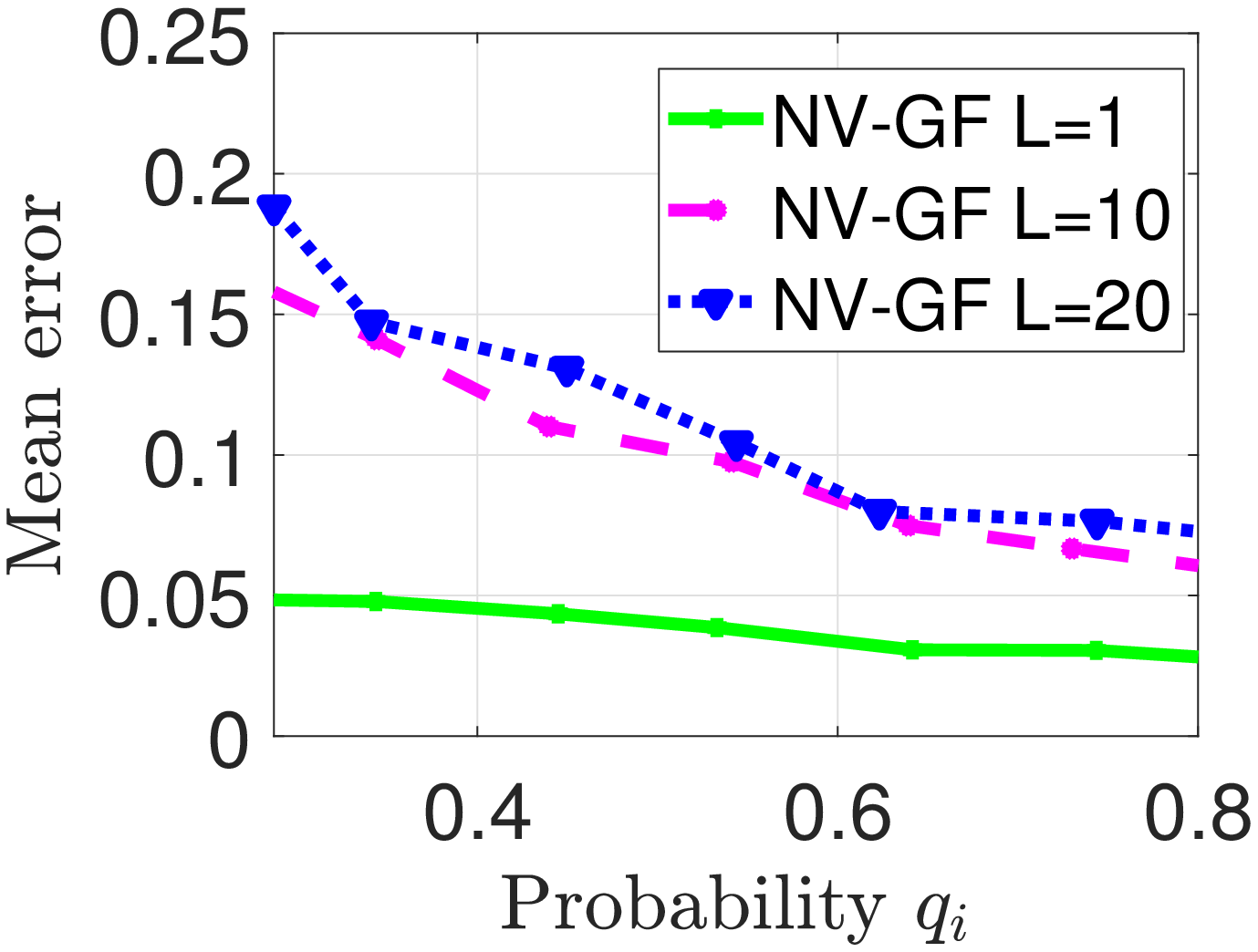}}\hspace{0.02cm}
\subfloat[${{\overline{\sigma}}_{\bold e}^{2}}$ with   increasing values of $q_{i}$ for  different  $L$ ]{\includegraphics[scale=0.32]{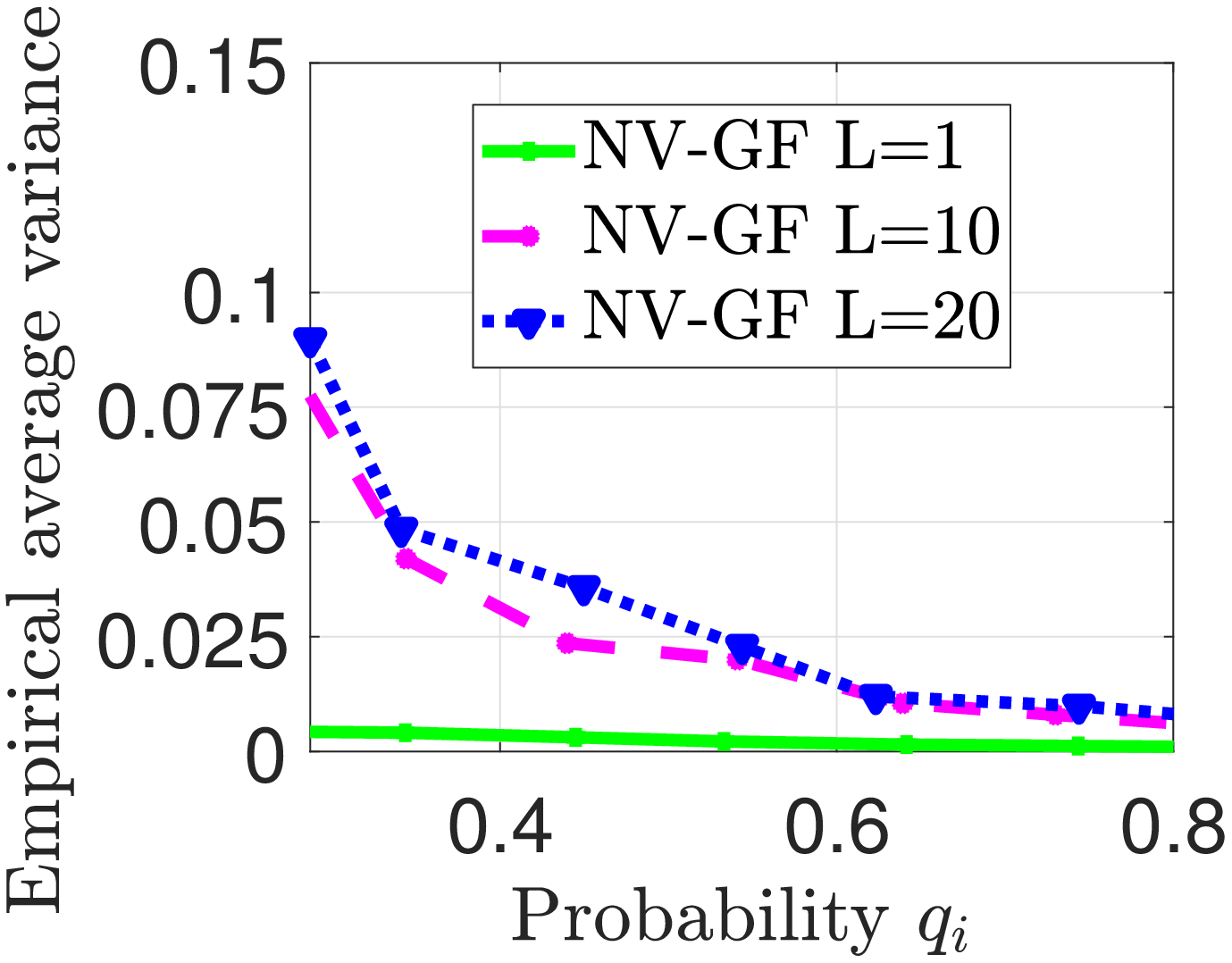}}\hspace{0.02cm}
\end{center}
\caption{The mean error  and the  variance ${{\overline{\sigma}}_{\bold e}^{2}}$ among all nodes and realizations
between the graph filtering  operated on the deterministic graph $\mathcal{G}_0$ and the time-varying graph $\mathcal{G}_t$,
for different types of  graph filters and orders $L$ of the filter. 
The  parameters are $N=100$, $ R_\text{B}=70$ m, $\mu=0.001$ and $\ell_s=150$ m.
} 
\label{fig-performance}
 \vspace{-0.2cm}
\end{figure*}


Initially, the states of the nodes
are  as follows:
$\mathcal{A}[0]=\emptyset$, 
$\mathcal{F}[0]=\emptyset$, 
$\mathcal{C}[0]=\mathcal{V}$. 
Let $n_{{t}_x}$  denote the current total number of nodes that  have been  allocated transmission slots (initially $n_{{t}_x}=0$) 
and   ${\xi}^{(\tau)}$  the set of  positions of the nodes allocated at slot $\tau$. 
As shown in Algorithm \ref{algorithm},  a single 
node $i$, randomly selected from the candidate set $\mathcal{C}[\tau]$, is  initially
activated to initiate the allocation of  transmitters at slot $\tau$, allowing thus to select  the   nodes  that can  transmit simultaneously with it at    the  same slot $\tau$.
First, the active node $i$  sends  to all other nodes its position  $pos_i$ and the current estimated  number of interfering nodes  $\hat{n}_I$,
which is computed from the  estimation of the total number of nodes as  $\hat{n}_I= \hat{N}-n_{{t}_x} - 1$. Then, the  states of nodes
  become  as follows:
$\mathcal{A}[\tau]=\{i \}$, 
$\mathcal{C}[\tau]=\mathcal{V}\setminus \{ \{i \} \cup  \mathcal{T}^{(\tau-1)}_a \cup  \ldots \cup  \mathcal{T}^{(0)}_a \}$, $\tau=1$.

At a certain slot $\tau$, every  node $u \in \mathcal{C}[\tau]$ computes the radius of the collision area $ R_\text{C}^{\hat{n}_I}$ by using  (\ref{equation_Rc}) and the received value 
 $\hat{n}_I$. Then, it computes its preventing range $R_P^{\hat{n}_I}$  by using (\ref{equation_Rp}).
In order to be able to satisfy    (\ref{condition-SINR}), the node $u$ must have 
 its preventing area of radius $R_P^{\hat{n}_I}$ outside
the preventing areas of the active node $i$ and of the other feasible nodes\footnote{Notice that, a candidate node can overhear some feasible nodes in its neighborhood but it {cannot} have the knowledge of all the feasible nodes.}
in $\mathcal{F}[\tau]$, as shown in Fig.~\ref{fig-allcation-slots}.
 We refer to the 
condition  where  the preventing area of a  candidate node $u$ is outside
the preventing area of the active node  as  \textit{Prevent-Condition}.
 Any  node $u$ checks if it  satisfies the \textit{Prevent-Condition} and if its preventing area is not overlapping
 with any potentially overhearing feasible nodes in its neighborhood. This overlapping occurs when the distance between two  nodes is less than  twice of the preventing radius.   
If both conditions are satisfied, the node $u$ sends a packet with its position $pos_u$ and its  state as potential \textit{feasible} node  to the active node $i$. 
As a result, the active node $i$  makes the   update $\mathcal{F}[\tau]=\mathcal{F}[\tau] \cup \{u \}$.
Note that the knowledge of the state of the nodes (i.e., $\mathcal{A}[\tau]$, $\mathcal{F}[\tau]$ and $\mathcal{C}[\tau]$) does not need to be known and shared among all the nodes.
Only the active node at a given time slot $\tau$ needs to keep  the set of feasible nodes that have notified to it.

After a predefined timeout\footnote{This maximum timeout can be estimated by taking into account the number of nodes and the delays related to transmission, reception and propagation.}
$t_{\text{out}}$ that ensures the reception of the packets from  all  potential feasible nodes, 
the active node $i$ checks if there are some nodes in  $\mathcal{F}[\tau]$ 
that have their corresponding  preventing areas  overlapping. 
This problem may happen when two or more
nodes change their states  at the same time instant, 
without hearing each other or when they are outside the transmission ranges of  each other.
The active node resolves the conflict\footnote{A simple way to resolve this conflict is to consider the order of receiving the feasible nodes by keeping adding as feasible node only the node that does not conflict with the previous selected feasible nodes. We leave
the extention to more advanced  approaches to solve this conflict for future research.}
by keeping only one of the conflicting feasible nodes. 
Next, in order to be able to decide which nodes will be selected as transmitters at 
slot $\tau$,
the active node $i$ compares the number of feasible nodes and the number of interfering nodes:

  $\bullet$ If $| \mathcal{F}[\tau]| \neq \hat{n}_I$, the active node $i$ decreases the  estimated  
number of interfering nodes  $\hat{n}_I=\hat{n}_I-1$ if $\hat{n}_I > 0$ 
and sends this new value to all  other  nodes that have not yet been allocated a slot. 
The process  is repeated by making the active node $i$ and the candidate nodes in $\mathcal{C}[\tau]$ update 
their preventing range $R_P^{\hat{n}_I}$ based on 
the new value of  $\hat{n}_I$
and checking again  if the \textit{Prevent-Condition} is satisfied.
Note that after this, the set  $\mathcal{F}[\tau]$ will also change.
Notice also that in order to reduce the overhead of these  control packets, 
every candidate node only needs  to send once a control packet informing about the potential feasibility        
to the same current active node, which has to
check for any new updated preventing area if  previous received feasible nodes are still not conflicting with each other and  satisfy the \textit{Prevent-Condition}.

 $\bullet$ If $| \mathcal{F}[\tau]| =\hat{n}_I$, the active node $i$ sends a packet to inform the candidate and feasible nodes
that   the nodes in  $\mathcal{F}[\tau]$  have been allocated the slot $\tau$  
$\big($i.e., $\mathcal{T}_{t_x}^{(\tau)}=\mathcal{F}[\tau]    \cup \{i\} \big)$, their positions are  ${\xi}^{(\tau)}$ and the total 
number of allocated
nodes   is now $n_{{t}_x}$, where
$n_{{t}_x}=n_{{t}_x}+| \mathcal{F}[\tau]|+1$.
Then, the process continues to determine the nodes that  will be allocated in the following 
slot  $\tau +1$, by 
 activating  randomly a new  node $i'$ from the remaining  candidate transmitter set 
 $\mathcal{C}[\tau+1]$, where
$\mathcal{C}[\tau+1]=\mathcal{C}[\tau] \setminus \mathcal{T}_{t_x}^{(\tau)}$,
$\mathcal{F}[\tau+1]=\emptyset$, $\mathcal{A}[\tau+1]=\{i' \}$.
The new active node $i'$ sets
the number of interfering nodes to 
$\hat{n}_I= \hat{N} - n_{{t}_x} -1 $
and informs  other nodes.
Then,  the process 
 is repeated by making every candidate node 
 determine
its new preventing range $R_P^{\hat{n}_I}$ based on 
the new value of  $\hat{n}_I$
and checking if it  satisfies the \textit{Prevent-Condition} and does not conflict with potentially
 overhearing  feasible nodes.

\begin{figure*}[ht!]

\begin{center}
\subfloat[${P}=0$ dBm, $\kappa=1$]{\includegraphics[scale=0.38]{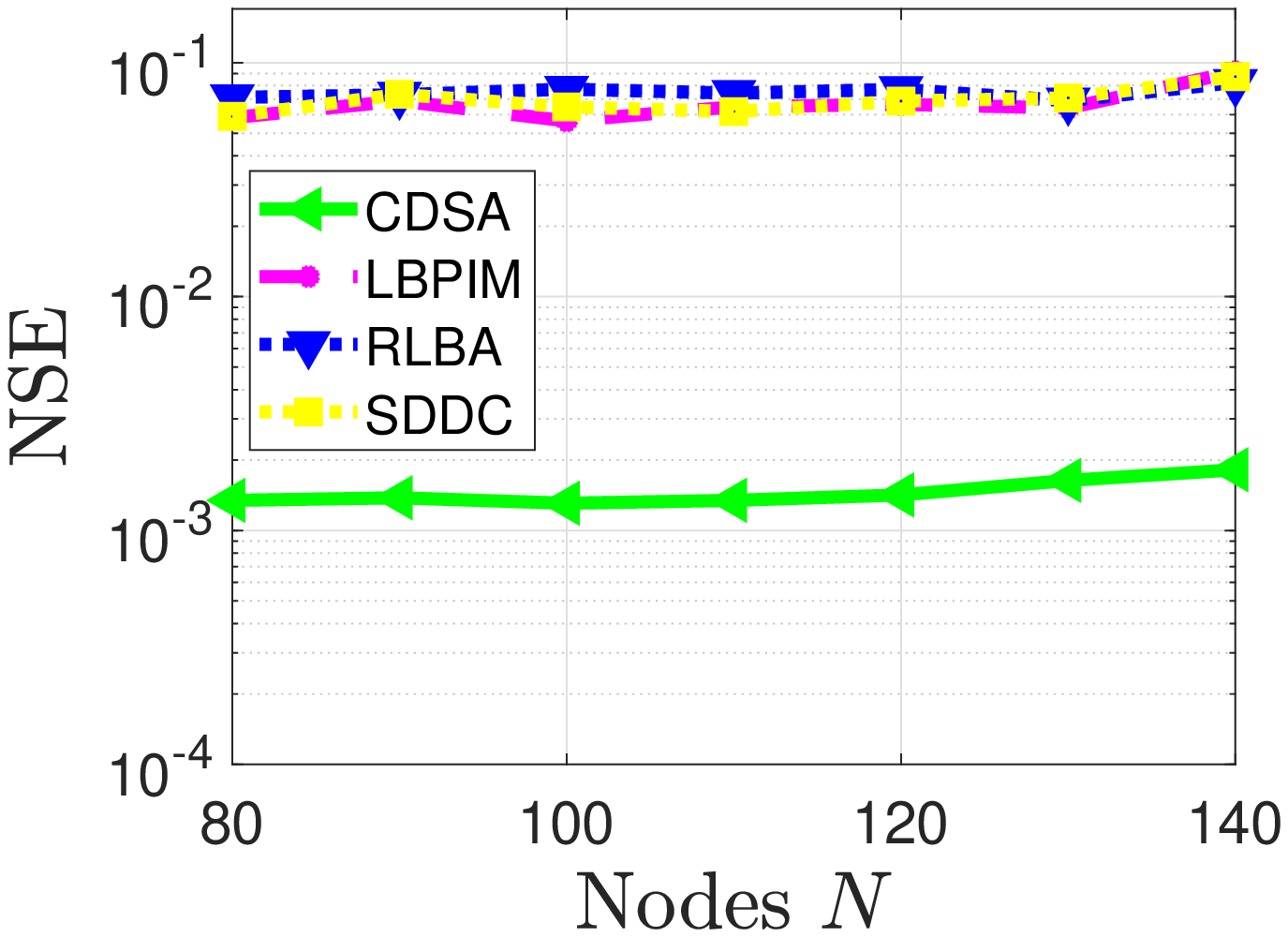}}\hspace{0.05cm}
\subfloat[$\kappa=1$, $N=100$]{\includegraphics[scale=0.38]{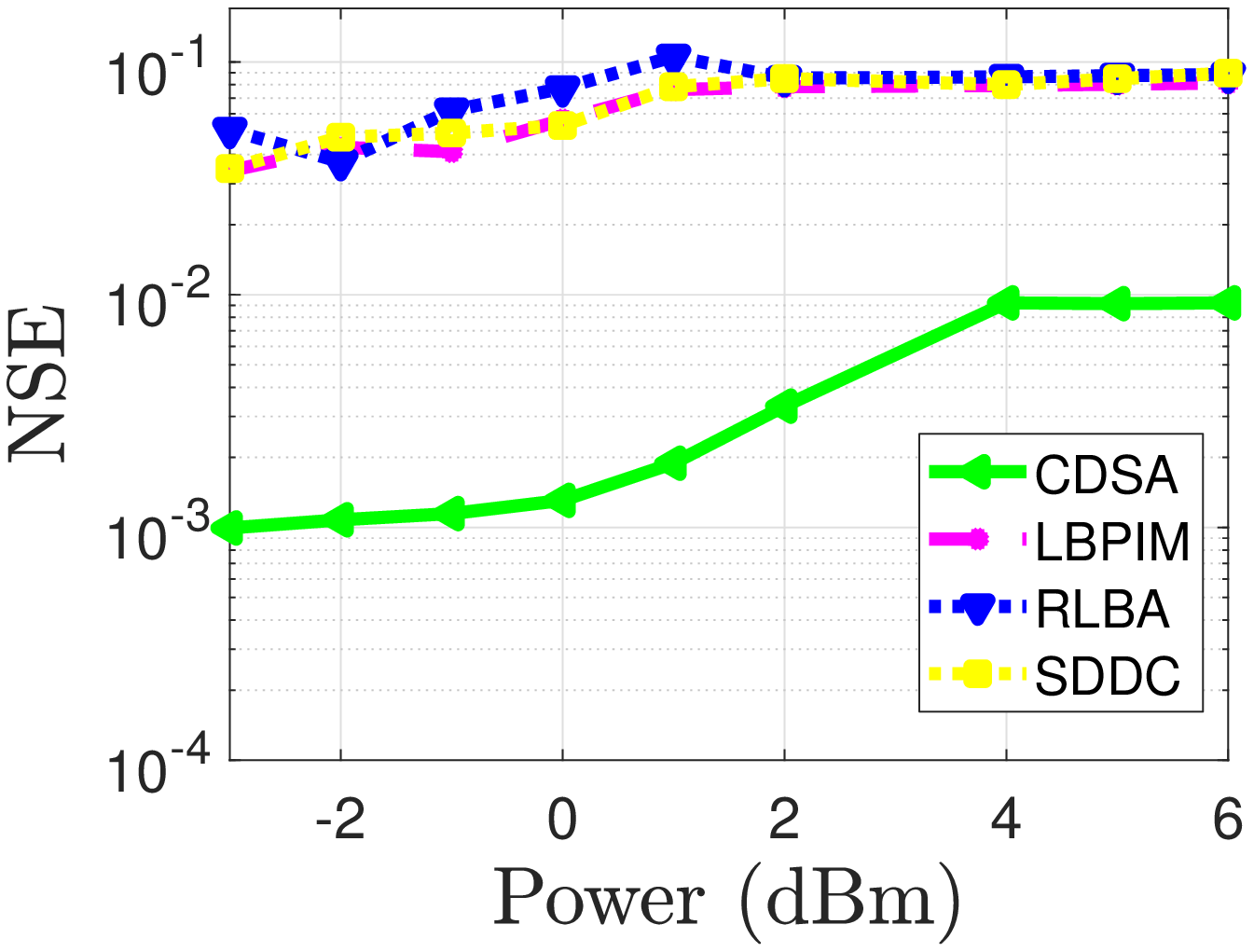}}\hspace{0.05cm}
\subfloat[${P}=0$ dBm, $N=100$]{\includegraphics[scale=0.38]{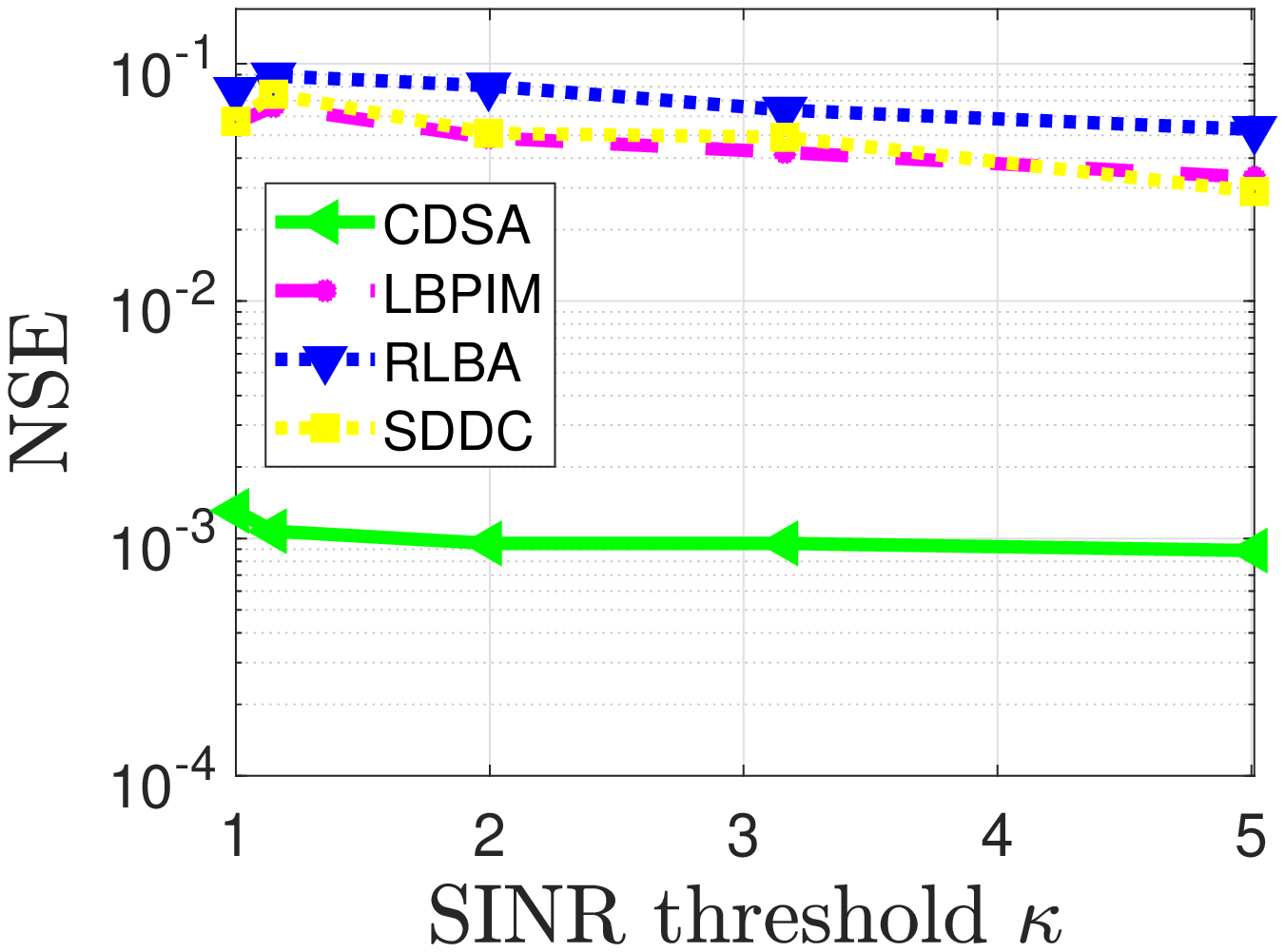}}
\end{center}
\caption{{NSE  between  the outputs corresponding to applying graph filtering  over  time-varying and static randomly deployed WSNs, when using
different  distributed scheduling algorithms at the MAC layer: CDSA, LBPIM~\cite{GoussevskaiaMoscibroda2008}, RLBA~\cite{YuHua2012} and SDDC \cite{FuchsPrutkin2015}. 
The  parameters are:  
$\mathscr{N}_o=-100$ dBm, 
$\nu=2.5$, $\chi=0.5$, $L=5$,  $z=176$ bits,  
$\ell_s=150$ m and $\mu=0.001$.}} 
\label{fig-performance-protocols}

\end{figure*}

\begin{figure}[ht!]
\begin{center}
\includegraphics[scale=0.4]{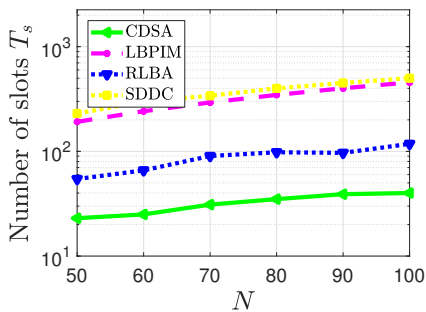}
\end{center}
\caption{{Number of slots $T_s$ (overall delay) needed for all nodes  to perform a successful broadcast with
different  distributed scheduling algorithms: CDSA, LBPIM~\cite{GoussevskaiaMoscibroda2008}, RLBA~\cite{YuHua2012} and  SDDC \cite{FuchsPrutkin2015}
when  applying graph filtering over  time-varying graphs. This corresponds to a graph filtering operation. The  parameters are:  
$\mathscr{N}_o=-100$ dBm, ${P}=-2$ dBm, $\kappa=1$,
$\nu=2.5$, $\chi=0.6$, $R_B=60$ m, 
$z=176$ bits, 
$\ell_s=280$ m and $\mu=0.001$.}
} 
\label{fig-delay}
 \vspace{-0.3cm}
\end{figure}

 $\bullet$ If  the estimated number of interfering nodes reaches zero ($\hat{n}_I = 0$), 
only the active node $i$ is assigned                   
the current  slot $\tau$ $\big($i.e., $\mathcal{T}_{t_x}^{(\tau)}= \{i\}\big)$ with $n_{{t}_x}=n_{{t}_x}+1$. Then, for 
the next slot $\tau+1$, we have
$\mathcal{C}[\tau+1]=\mathcal{C}[\tau] \setminus \{i\}$, $\mathcal{F}[\tau+1]=\emptyset$.
The process is repeated so that   a new candidate node from $\mathcal{C}[\tau+1]$ is randomly chosen as active node. This implies
setting  $\hat{n}_I= \hat{N} - n_{{t}_x} -1 $
and checking which
 candidate nodes satisfy the  \textit{Prevent-Condition}
  and do not conflict with potentially
 overhearing feasible nodes.

The algorithm stops whenever there is 
no node  that needs to determine its allocation slot, which means that
the candidate  set 
becomes empty $\mathcal{C}[\tau]=\emptyset$.
At the end of the  algorithm,
every  transmitter  $i$ knows  its assigned slot $\tau$, the number of other nodes  that will transmit at the same time slot $\tau$
and their positions ${\xi}^{(\tau)}$. 
Thus, every transmitter  $i$ is able to compute the smallest $\text{SINR}$  in its neighborhood, given by:
 \vspace{-0.1cm}
 \begin{equation}
   {\text{SINR}}^{\text{min}}_i= \min_{j  \in \Delta^B_i} \left(\frac{ \; \displaystyle \frac{{P}}{ d_{i,j}^{\nu}}} { \displaystyle \sum_{u \in \mathcal{T}_{t_x}^{(\tau)},\; u \neq i }^{}    \; \; \frac{{P}}{ d_{u,j}^{\nu}}  +\mathscr{N}_0}  \right), \;   i \in \mathcal{T}_{t_x}^{(\tau)} 
  \label{SINR-min}
 \end{equation}

 \noindent {Then, the main idea is that each transmitter $i$ can adjust the PDRs of its neighbors within its broadcast area to make them equal, which is 
necessary to maximize the accuracy of the graph filtering process, as shown in Section \ref{GF_symLinks},
 by setting them to  $\text{PDR}^{\text{min}}_{i}$, which refers to
 the minimum PDR of node $i$ when broadcasting to its neighbors within the range $R_\text{B}$.}  
 Notice that $\text{PDR}^{\text{min}}_{i}$ can be readily obtained from the $\text{SINR}^{\text{min}}_i$  estimated in the neighborhood of node $i$,
 by using (\ref{eq-ber}) and (\ref{eq-pdr}), as follows:
  \begin{equation}
 \text{PDR}^{\text{min}}_{i}=  \bigg(1 -   \varsigma_1 \displaystyle \sum^{\varsigma_2}_{k=2} (-1)^k  e^{\varsigma_3 \; \text{SINR}^{\text{min}}_i \; \big(\frac{1}{k}-1\big)}  \bigg)^z 
 \label{eq-pdr-min}
\end{equation}

 \noindent In order to  equalize the PDRs, each transmitter $i$ 
 broadcasts a packet to all its neighbors $j  \in \Delta^B_i$, but imposes for  each neighbor $j$ a different
    probability of acceptance  $p^{(\text{ac})}_{ij}$ to accept the reception of the packet, and which is chosen such that:
 \begin{equation}
  \text{PDR}^{\text{min}}_{i}   =  p^{(\text{ac})}_{ij} \; \text{PDR}_{ij},  \;  i \in \mathcal{V}, \;{\forall    j \in  \Delta^B_i}.
   \label{eq-pac}
 \end{equation}

\noindent In other words, for a given  probability $p^{(\text{ac})}_{ij}$, 
the transmitter $i$, during its allocated slot $\tau$, specifies in every  packet it  broadcasts,  the  identifier of the 
 nodes that have to accept   the received packet with proportions that match $p^{(\text{ac})}_{ij}$. Other nodes simply ignore the  packet. 
 It is important to notice that this is due to the fact that our protocol is designed for ensuring accuracy in the filtering operations,
 as opposed to the case of maximizing throughput of  bits. 
 

\noindent The use of the probability $p^{(\text{ac})}_{ij}$ leads to  the adjusted values of PDRs that allow to determine
 the desired connection
probability matrix $\bold Q$, as shown in Section \ref{GF_symLinks}.  Then, the filter coefficients can be easily optimized
by using  (\ref{eq-minimizing-MSE}). 
As our experimental results show  in Section \ref{sec_experim}, each node can perform the distributed graph filtering task with  high accuracy due to the  control of the
resulting bias and variance of the graph filtering process.

As mentioned earlier, in order to maximize the time efficiency in our graph filtering, our scheduling protocol aims at minimizing
the total number of slots. The main result is given by the
following proposition. 

\begin{proposition}
Consider $N$ sensor nodes  deployed  uniformly random inside a 2-D  square area of side length $\ell_s$.
 In order to  decrease the number of  allocated slots in our CDSA protocol, thus   allowing a higher number of simultaneous transmissions,
 the probability of having all  nodes inside a disc of
radius $R^{\ast}_P$ has to be
reduced by increasing the probability
 that there are any two nodes located at a distance higher
 than $2\;R^{\ast}_P$. This can be achieved if 
 $\ell_s$ is
 selected such as  $\ell_s >> R^{\ast}_P$ and reducing the value of $\chi$ while still maintaining  the connectivity of the whole network ($\ell_s \sqrt{(\pi N R^2_\text{m})^{-1} \log{N} }<\chi<1$),
 where $R^{\ast}_{P}$ is  given by:
 \vspace{-0.15cm}
$$R^{\ast}_{P}=R_P^{\hat{n}_I=1}=\chi \; \bigg( \frac{{P}}{\kappa \;\mathscr{N}_0 }  \bigg)^{ \frac{1}{\nu}} + \bigg( \frac{ \;\kappa \;{P}\;\;  R_\text{B}^{\nu} }{ {P}   - \kappa  R_\text{B}^{\nu} \;\mathscr{N}_o   }   \bigg )^{\frac{1}{\nu}}$$
 \end{proposition}
 
\textit{Proof}: See Appendix. 

 \begin{figure*}[ht!]
\begin{center}
\subfloat[Deterministic WSN ]{\includegraphics[scale=0.34]{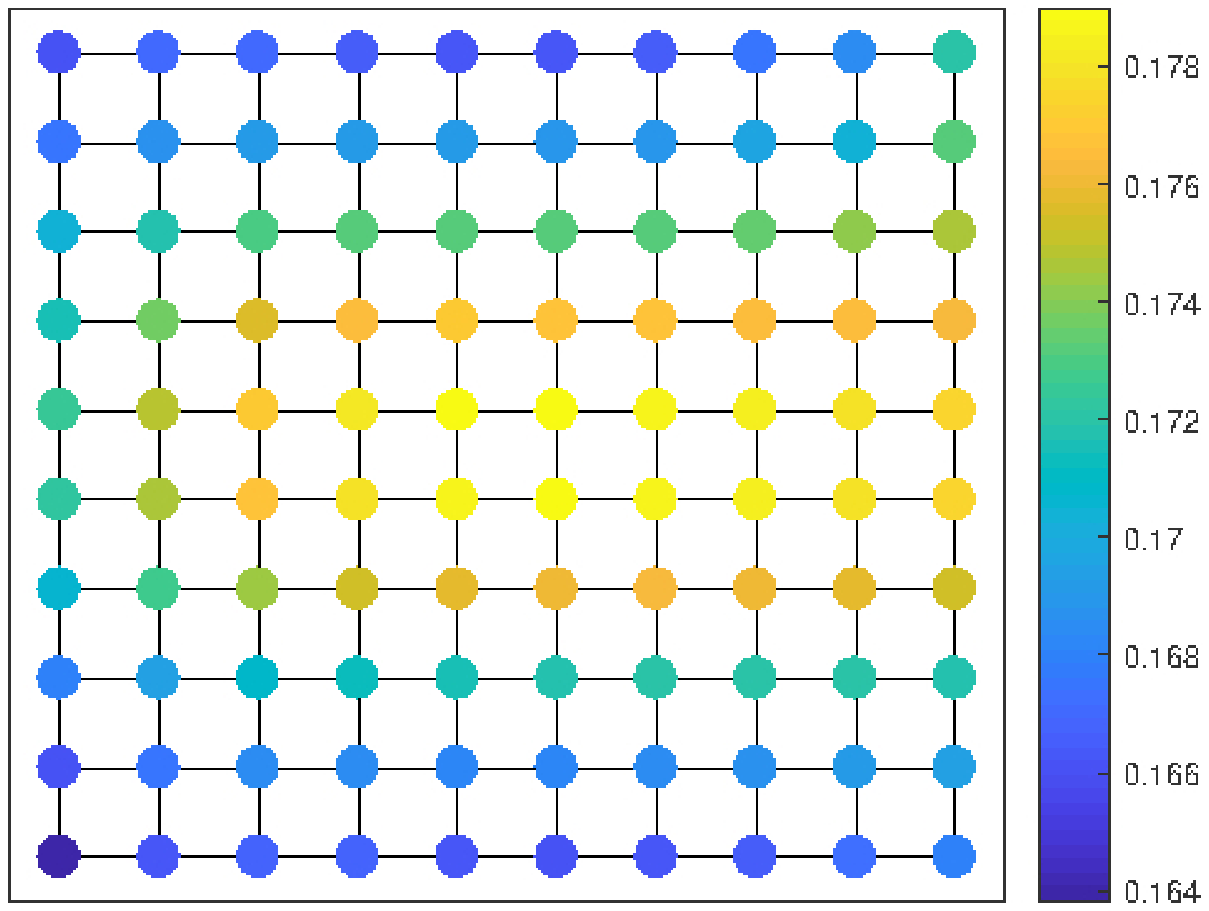}}
\subfloat[Time-varying WSN \& RLBA]{\includegraphics[scale=0.34]{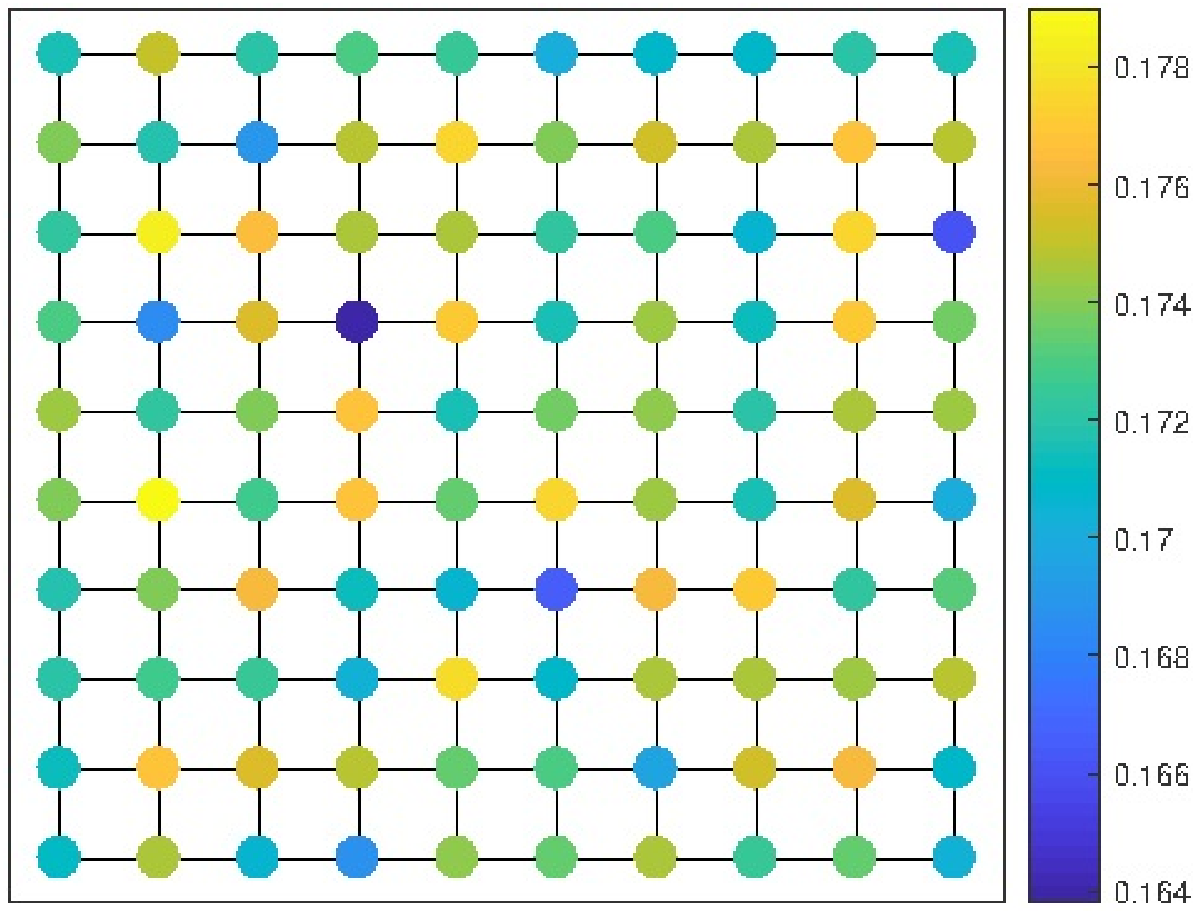}}
\subfloat[Time-varying WSN  \&  LBPIM]{\includegraphics[scale=0.34]{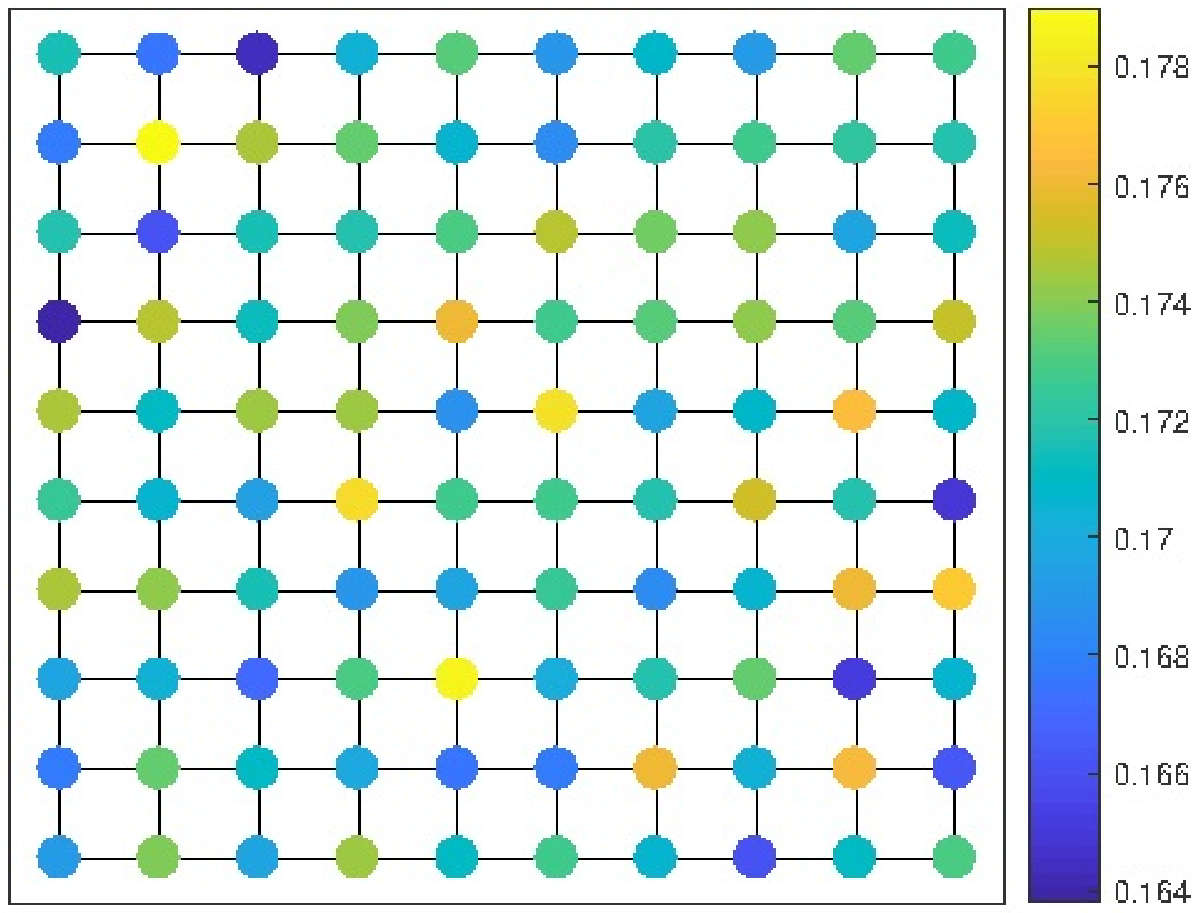}}
\subfloat[Time-varying WSN \&  CDSA]{\includegraphics[scale=0.34]{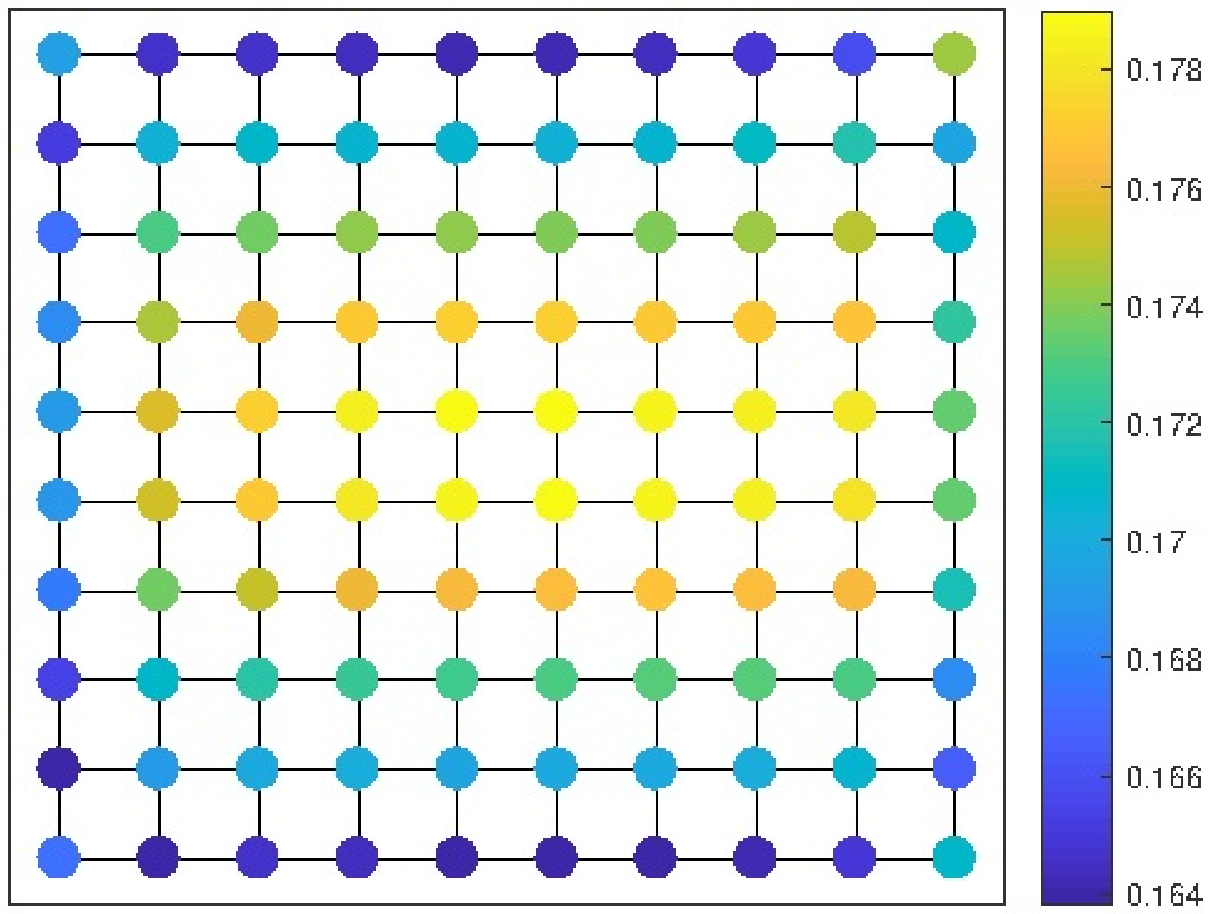}}\\
\subfloat[Deterministic WSN ]{\includegraphics[scale=0.34]{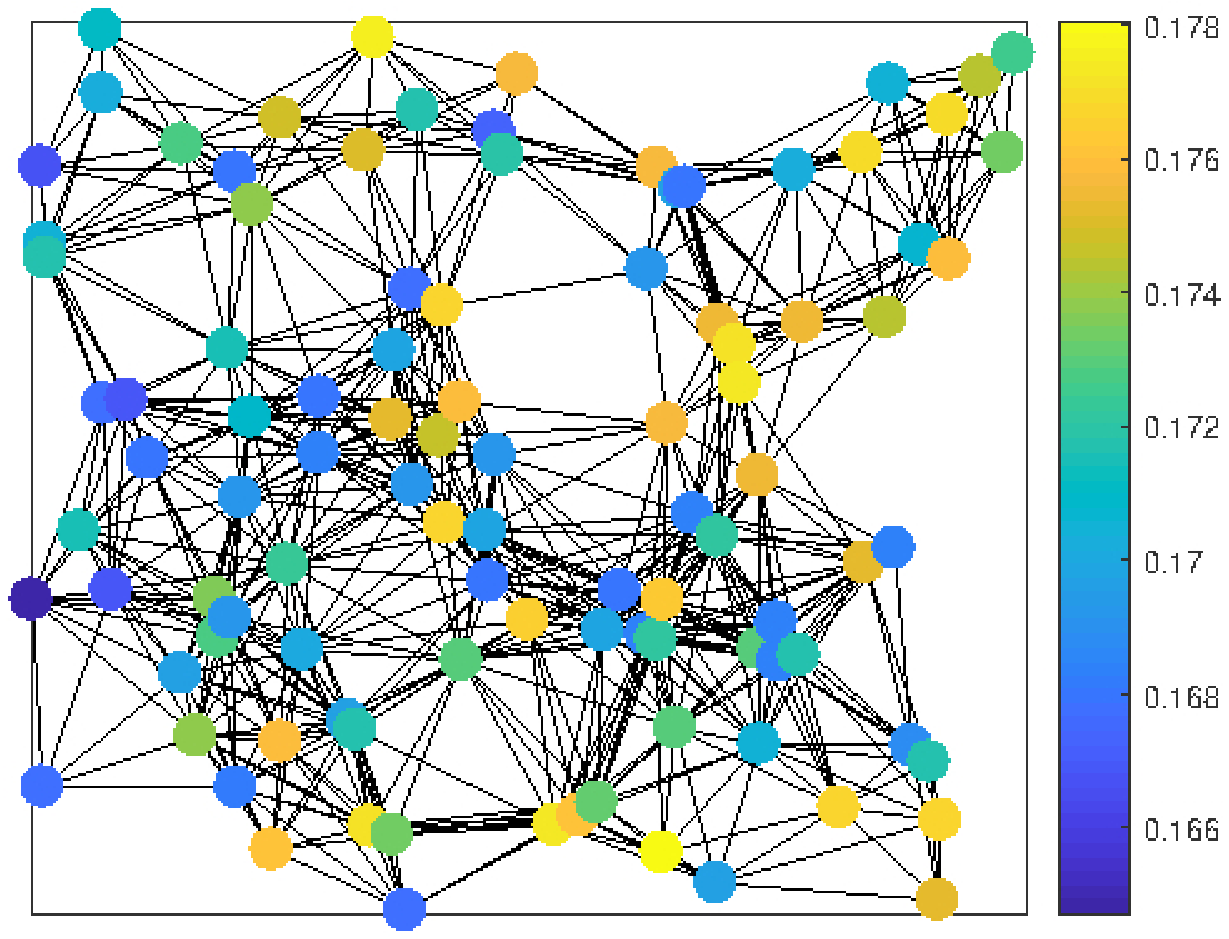}}
\subfloat[Time-varying WSN \& RLBA]{\includegraphics[scale=0.34]{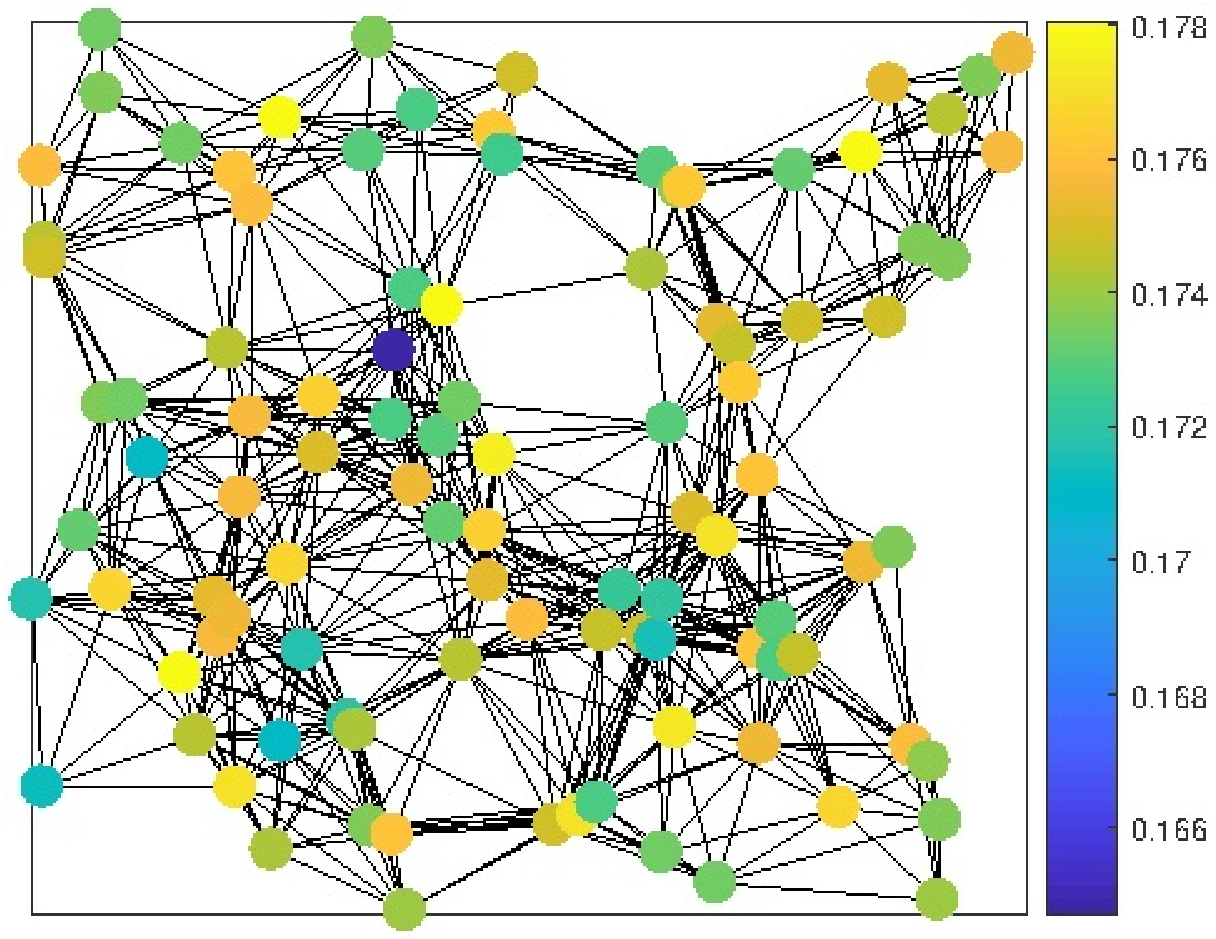}}
\subfloat[Time-varying WSN  \&  LBPIM]{\includegraphics[scale=0.34]{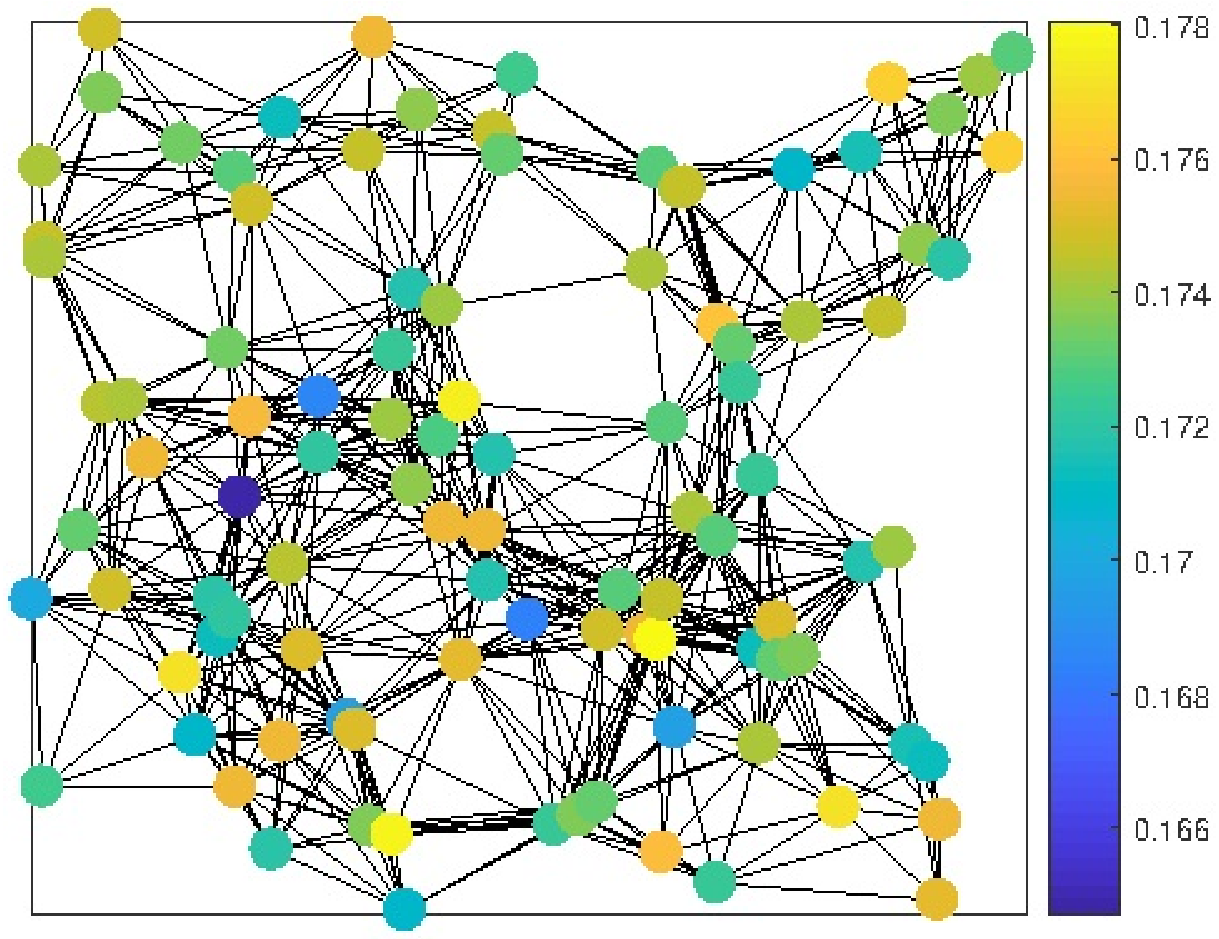}}
\subfloat[Time-varying WSN \&  CDSA]{\includegraphics[scale=0.34]{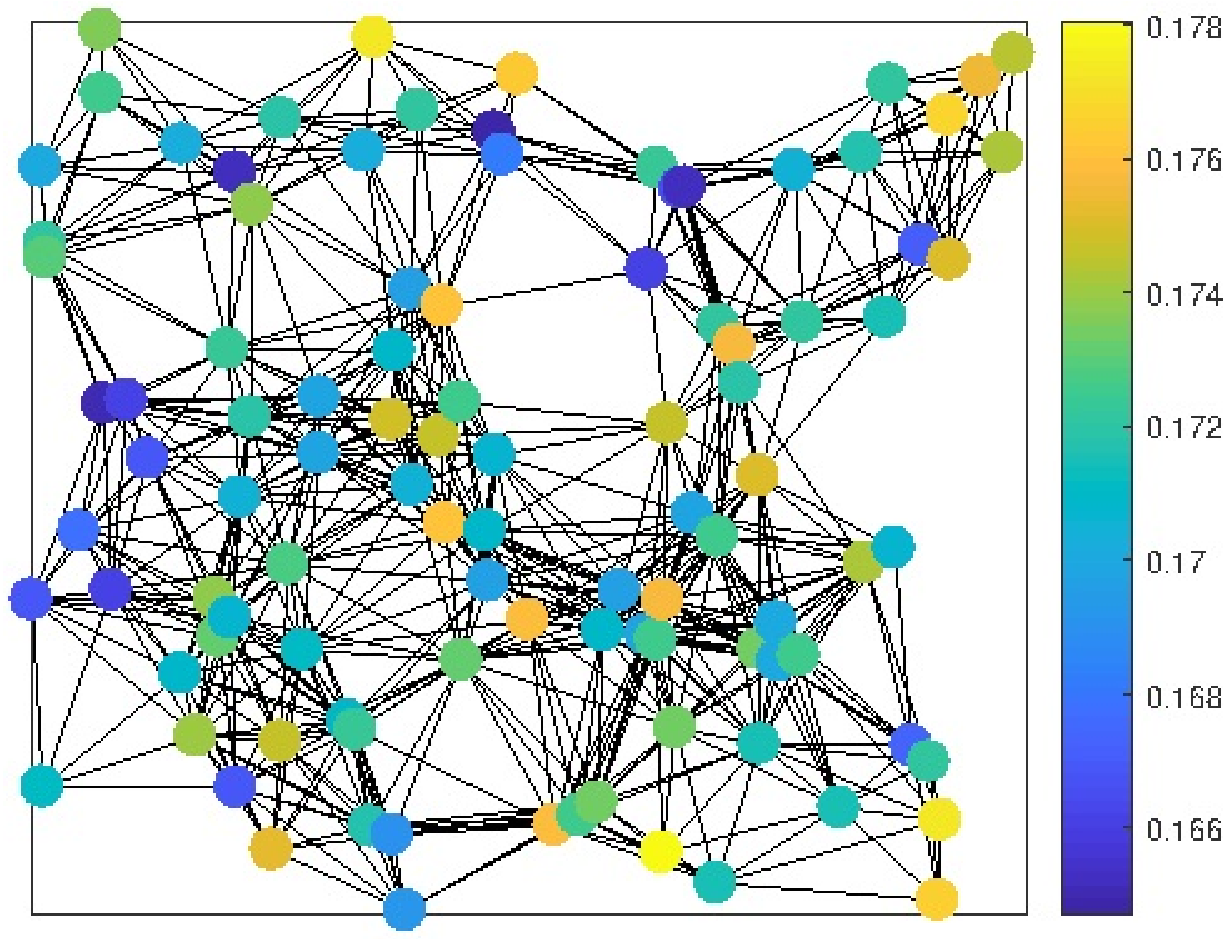}}\\
\end{center}
\caption{{Example of   denoising by graph filtering   in a $10{\times}10$-grid  and random WSNs; 
(a) and (e) correspond to graph signal outputs  in the static deterministic WSN with perfect MAC environment. (b), (c), (d),  (f), (g) and (h)  average graph signal outputs in  time-varying WSNs
for different scheduling algorithms. 
The color of each node depends on the  signal  value  at that node. The  parameters are:  
${P}=-2$ dBm, $\kappa=1$, $\mathscr{N}_o=-100$ dBm,
$\nu=2.5$, $\chi=0.5$,  $R_B=50$ m, $L=20$, $z=176$ bits, $w=0.45$, 
$250$ m $\leq \ell_s \leq 500$ m, $N=100$, $\mu=0.001$. Notice the higher closeness between (a) and (d) as well as (e) and (h).}}
\label{fig-denoising}
\vspace{-0.25cm}
\end{figure*}

\begin{remark}
{The proposed CDSA algorithm is
designed for sensor nodes that  perform many iterations of graph filtering process.
In order to reduce the control communication, these nodes can keep their
allocated slots in several iterations of the different graph filtering operations even though there are some link losses.
This is because the graph filters are designed to be robust to network topology changes, through the optimization
of the filter coefficients.}
\end{remark}

\vspace{-0.15cm}
\section{Numerical experiments}\label{sec_experim}

 {This section validates our theoretical findings, where several 
experiments are conducted in Matlab to evaluate the performance of our proposed solutions.
WSNs with $N$ sensor nodes  are
randomly and uniformly distributed over a square area of side $\ell_s$ m. Each node can communicate with the neighbors within
its broadcast range  given by $R_B=\chi \left( P /(\kappa \;\mathscr{N}_0) \right)^{ \frac{1}{\nu}}$.}
We consider the   input graph signal, acquired by the WSN,   given by $\bold x=\bold v+\bold n$,   
where  $\bold v$ is the smooth true graph  signal and $\bold n$  is a zero mean Gaussian noise with $0.1$ standard deviation. 
In order to  impose a small spectral norm
 that can further decrease the variance, the shift operator  used is 
 $\bold S=\lambda_{\max}^{-1} \bold L- 0.5 \bold I$, 
 where $\lambda_{\max}$ is the maximum eigenvalue of $\bold L$.  
 We analyze
the error $\bold e=\bold y_t{-}\bold y$ and 
the empirical  variance of the error averaged over all nodes and
realizations, that is, 
${{\overline{\sigma}}_{\bold e}^{2}}=\text{tr}(\mathbb{E}[\bold e \bold e^H])/N$,  which  can approach 
the average variance $\overline{var}  [\bold y_t]$ for a sufficiently high number of realizations.
The  filter coefficients used when operating filtering over time-varying networks are optimized as presented in Section
  \ref{minimizing-MSE}. 
The results are obtained by  averaging over 1000 realizations of graphs.

Fig.~\ref{fig-performance} plots 
the  mean error averaged over all the nodes and realizations,  as well as the empirical average variance ${{\overline{\sigma}}_{\bold e}^{2}}$,  
for  different  probabilities   $q_{i}$ of  link activation. 
As expected, it can be seen in Fig.~\ref{fig-performance}(a)-(b)    that 
  node-variant GF has significantly better performance than   the node-invariant GF,
 where in both filters  the coefficients are optimized. This is due to the higher number of degrees of freedom 
 that node-variant GF can offer to choose the coefficients, as compared to the node-invariant GF, where the same coefficients are used by all nodes.
In  Fig.~\ref{fig-performance}(c)-(d), we can observe that  by using node-variant GFs with optimized filter coefficients,
  a small mean error in the order of   $10^{-2}$ and a low empirical average variance in the order of   $10^{-3}$ are obtained  for 
$q_{i} \geq 0.55$.  This also means that  
the  output of  graph filtering over the deterministic graph $\mathcal{G}_0$ is very close to  the one obtained on average over the time-varying graph $\mathcal{G}_t$, indicating 
 that a high   filtering accuracy is achieved.
As expected, it can also be noticed that  better link connectivities (i.e., higher $q_{i}$'s) lead to higher graph filtering accuracy.

{We  evaluate also the performance of applying   graph filtering in random WSNs, by
comparing at MAC layer, our proposed CDSA protocol
with the three    
 state-of-the-art algorithms that allow asymmetric links, namely 
LBPIM \cite{GoussevskaiaMoscibroda2008},  RLBA \cite{YuHua2012} and SDDC \cite{FuchsPrutkin2015}}. 
{Fig. \ref{fig-performance-protocols}(a) shows that our  proposed  protocol CDSA significantly outperforms the existing algorithms
in term of the resulting Normalized  Squared Error $\text{NSE}= \| \bold y -  \bar{\bold y}_t \|^2 /\| \bold y \|^2$  
of the graph filtering. }
In fact, compared to the
existing algorithms, our CDSA protocol controls  the PDRs, which ensures the accuracy of the filtering operations. 
%
%
Fig.~\ref{fig-performance-protocols}(b) shows that  
 increasing the transmission power, increases  
the NSE, which is caused by the fact that more errors are generated due to 
involving more nodes inside a larger broadcast area, where the graph filtering tasks are performed.  
Nevertheless, our CDSA protocol still achieves the highest filtering accuracy,
which  can be even   reached in a single filtering iteration.
 Fig.~\ref{fig-performance-protocols}(c) shows the impact of the  threshold $\kappa$
on the NSE of graph filtering. Indeed, increasing the threshold $\kappa$  improves  the SINR at 
the receivers as well as  the PDRs, which reduces  the resulting filtering error.


{Fig. \ref{fig-delay} illustrates  the number of time slots $T_s$ until  all nodes   perform a successful local broadcast per  graph filtering iteration,
which accounts for the total delay per graph filtering operation. 
The results show that our CSDA algorithm achieves lower delay compared to the three other protocols.} 
This can be explained by the fact that
our CSDA protocol controls the PDRs and
takes into account 
the number of interfering nodes and their locations to determine the
preventing area when allocating the slots, ensuring that the $\text{SINR}$ at all receivers is always higher than $\kappa$. 
This approach is different from the ones used by the three other protocols, 
where each node can transmit with a certain probability in each slot, without considering the number of interfering nodes.


Fig. \ref{fig-denoising} shows two examples  of denoising by graph filtering in two different topologies, namely, a $10{\times}10$-grid   and random WSNs, 
where    different
scheduling algorithms are used. The  average graph signal output obtained by graph filtering in  time-varying WSNs 
when using CDSA protocol, is the  one that 
matches better the  graph signal output obtained by graph filtering in the   deterministic WSN with perfect MAC environment
(i.e.,  absence of interference and background noise). 
This is due to the fact that  our CDSA protocol  imposes during the slot allocation a preventing area  that takes into account the number of
interfering nodes and controls the PDRs at each broadcast region to improve the graph filtering accuracy. 

 \vspace{-0.15cm}
\section{Conclusion}
In this work, we first  study and characterize  the graph  filtering error and show  that  for  both types of FIR GFs (node-invariant and node-variant),
 equal probabilities for all the links,  enables to have an unbiased filtering, which {cannot} be achieved in practice in  WSNs due to interferences and noise.
Then, we present an efficient and robust design strategy  to perform  graph
filtering tasks over random WSNs with  node-variant graph filters
by maximizing 
accuracy that is, optimizing a  bias-variance tradeoff.
 The accuracy and the efficiency of the graph
filtering process, which is implemented distributedly by means 
of cooperation and communication exchanges between the sensor nodes, is enforced  at the MAC layer by designing
a Cross-layer Distributed Scheduling Algorithm.
As illustrated in the experiments, a high filtering accuracy is obtained 
when our proposed CDSA protocol combined with optimized graph filtering coefficients  is used, allowing to
obtain higher performance for the  denoising application, as compared
to the best existing state-of-art algorithms.







  \appendix
  \section*{}
  \vspace{-0.25cm}
 \subsection*{Conditions for obtaining an unbiased graph filtering}
In order to obtain an unbiased filtering i.e., $\bar{\bold e}=\bold 0$, we must have:
\vspace{-0.25cm}
\begin{equation}
\displaystyle\sum_{l=0}^{L} \phi_{l}   \; \bar{\bold S}^{l} \; \bold x=\displaystyle\sum_{l=0}^{L} h_{l}   \; \bold S^{l} \; \bold x
\end{equation}
\vspace{-0.15cm}

\noindent which is  equivalent to enforce the following:
\begin{equation}
 \phi_{0}  =h_{0}  \;\;\;  \text{and} \;\; \displaystyle\sum_{l=1}^{L} {\big({\phi_{l} }^{\frac{1}{l}}    \; \bar{\bold S}\big)}^{l} \; \bold x=\displaystyle\sum_{l=1}^{L} \big( {h_{l}^{\frac{1}{l}}   \; \bold S}\big)^{l} \; \bold x
\end{equation}
\vspace{-0.15cm}

\noindent If we select $\phi_{0}  =h_{0}$ and ${\phi_{l} }^{\frac{1}{l}}    \; \bar{\bold S}={h_{l}^{\frac{1}{l}}   \; \bold S}$ for $l \geq 1$,
we have  $\bar{\bold e}=\bold 0$, implying that we can write the following conditions:
\begin{align}
{\phi_{l} }^{\frac{1}{l}}    \bar{\bold S}&=\begin{bmatrix}
{\phi_{l} }^{\frac{1}{l}} p_{11}s_{11} & {\phi_{l}}^{\frac{1}{l}} p_{12}s_{12}& \cdots  & {\phi_{l}}^{\frac{1}{l}} p_{1N}s_{1N} \hspace{-0.2cm}\\
{\phi_{l} }^{\frac{1}{l}} p_{21} s_{21} & {\phi_{l} }^{\frac{1}{l}} p_{22} s_{22} & \cdots   & {\phi_{l} }^{\frac{1}{l}} p_{2N}s_{2N}\hspace{-0.2cm}\\
\vdots & \vdots & \vdots &  \vdots &\hspace{-0.2cm}\\
{\phi_{l} }^{\frac{1}{l}} p_{N1} s_{N1} & {\phi_{l} }^{\frac{1}{l}} p_{N2} s_{N2}&   \cdots & {\phi_{l}}^{\frac{1}{l}} p_{NN} s_{NN}\hspace{-0.2cm}
\end{bmatrix}
\end{align}
\begin{align*}
&={h_{l}^{\frac{1}{l}}   \bold S}=\begin{bmatrix}
{h_{l}}^{\frac{1}{l}}  s_{11} & {h_{l}}^{\frac{1}{l}}  s_{12}& \cdots  & {h_{l} }^{\frac{1}{l}} s_{1N}\\
 {h_{l}}^{\frac{1}{l}}  s_{21} & {h_{l} }^{\frac{1}{l}}  s_{22} & \cdots  & {h_{l} }^{\frac{1}{l}} s_{2N}\\
\vdots  & \vdots & \vdots  &  \vdots &\\
 {h_{l}}^{\frac{1}{l}}  s_{N1} & {h_{l}}^{\frac{1}{l}}   s_{N2}&  \cdots  & {h_{l}}^{\frac{1}{l}}   s_{NN}\end{bmatrix}\\ \nonumber
\vspace{-0.15cm}
 \end{align*}
\noindent which means that we need  to impose the following conditions: 
\begin{align*}
{\phi_{l} }^{\frac{1}{l}} p_{ii} s_{ii} & = {h_{l}}^{\frac{1}{l}} s_{ii}  \\
{\phi_{l} }^{\frac{1}{l}} p_{ij} s_{ij} & = {h_{l}}^{\frac{1}{l}} s_{ij},  \;\; \;\;  \; \; \forall i,j,   \;\;  l \geq 1\\
{\phi_{l} }^{\frac{1}{l}} p_{ji} s_{ji} & = {h_{l}}^{\frac{1}{l}} s_{ji} \\
\end{align*}
\vspace{-0.15cm}

\vspace{-0.7cm}
Therefore, it can be easily seen that to obtain  $\bar{\bold e}=\bold 0$, 
the links need to be activated with an equal probability $p$ and
the  coefficients have to meet the following requirements:
\begin{equation}
\phi_{l}   = p_{ij}^{-l}  \;\;{h_{l}} = p_{ji}^{-l} \;\;{h_{l}}= p_{ii}^{-l} \;\;{h_{l}}={p}^{-l}\; h_{l} \;\; \forall i,j,l
\end{equation}

 \vspace{-0.25cm}
\subsection*{Proof of \textbf{Proposition 1}}

\noindent By considering  the linearity of expectation and trace, the first term on the right side of (\ref{eq_variance}) can be written as \cite{IsufiLoukasSimonetto2017}:
\vspace{-0.35cm}

\begin{small}
\begin{equation}
\begin{split}
 \text{tr}(\mathbb{E} \big[\bold y_t {\bold y_t}^{H}])  & = {\!\sum^{L}_{k=0, l=0}} \Lambda(k,l)
\end{split}
\vspace{-0.25cm}
\end{equation}
\end{small}
\vspace{-0.25cm}
\noindent where: 
\vspace{-0.25cm}
\begin{small}
\begin{equation}
\begin{split}
 \Lambda(k,l) =  & \text{tr}\bigg(\mathbb{E}\bigg[  \text{diag}(\boldsymbol \phi^{(k)}) \bold \Theta(t,t-k+1) \; \bold x    \bold x^H \\
 & \times {\bold \Theta(t,t-l+1)}^H{{\text{diag}(\boldsymbol \phi^{(l)})}^H}\bigg]\bigg).  
\end{split}
\vspace{-0.25cm}
 \end{equation}
\end{small}
\vspace{-0.3cm}


\noindent By considering the fact that the trace is commutative with  respect to the expectation and 
 invariant under cyclic permutations i.e.,
$\text{tr}(\bold M \bold W \bold Z)=\text{tr}(\bold Z \bold M \bold W)$,  we can write:
\vspace{-0.15cm}

\begin{footnotesize}
\begin{equation}
\begin{split}
&\Lambda(k,l)=\mathbb{E}\bigg[ tr\big(\;   \text{diag}(\boldsymbol \phi^{(k)}) \;\bold \Theta(t,t-k+1) \;  \bold x   \bold x^H   {\bold \Theta(t,t-l+1)}^H \\
 & \times {diag(\boldsymbol \phi^{(l)})}^H     \big)\bigg] \\
 &{=}\text{tr}\big(\mathbb{E}\big[{\bold \Theta(t,t-l+1)}^H {\text{diag}(\boldsymbol \phi^{(l)})}^H  \text{diag}(\boldsymbol \phi^{(k)}) \;\bold \Theta(t,t-k+1)  \big]     \bold x    \bold x^H       \big)\\
 \end{split}\label{equa-commutativity}
 \vspace{-0.25cm}
 \end{equation}
\end{footnotesize}
\vspace{-0.15cm}


%

\noindent By making the observation that   for any square matrix $\bold M$  
and a positive semi-definite  matrix $\bold W$, the 
inequality $\text{tr}(\bold M \bold W)\leq  \| \bold M\|_2 \;\text{tr}(\bold W)$ \cite{SaniukRhodes87} holds,
and then applying it to (\ref{equa-commutativity}), we can write:
\vspace{-0.2cm}


\begin{small}
\begin{equation}
\begin{split}
&\Lambda(k,l){\leq}\big\| \mathbb{E}\big[ {\bold \Theta(t,t-l+1)}^H{{\text{diag}(\boldsymbol \phi^{(l)})}^H}\text{diag}(\boldsymbol \phi^{(k)})\\
 &  \times  \bold \Theta(t,t-k+1)\big]\big\|_2 \;tr \big( \bold x    \bold x^H       \big) 
   \end{split}
 \end{equation}
\end{small}
\vspace{-0.25cm}

\noindent By applying the Jensen's inequality of the spectral norm  \begin{small} $\| \mathbb{E}[\bold M] \|_2 \leq \mathbb{E}[ \| \bold M \|_2  ]$\end{small}
and the sub-multiplicativity property   of the spectral norm  of a square matrix \begin{small}$\| \bold M \bold W  \|_2  \leq \| \bold M   \|_2  \|  \bold W  \|_2 $\end{small}, we obtain:
\vspace{-0.25cm}
 
 \begin{footnotesize}
 \begin{equation}
\begin{split}
    &\Lambda(k,l){\leq}\mathbb{E}\big[ \big\| \;   {\bold \Theta(t,t-l+1)}^H {\text{diag}(\boldsymbol \phi^{(l)})}^H  \text{diag}(\boldsymbol \phi^{(k)}) \;\bold \Theta(t,t-k+1)   \big\|_2  \big] \\
    &\times tr \big( \bold x \;   \bold x^H       \big) \\
      &{\leq} \mathbb{E}\bigg[     \big \| \big(\prod_{\tau=t}^{t-l+1} \! \! \bold S_{\tau} \big)^H \big \|_2   {\| {{\text{diag}(\boldsymbol \phi^{(l)})}^H}\|}_2   {\| \text{diag}(\boldsymbol \phi^{(k)})\|}_2  \; \big \| \big(\prod_{\tau=t}^{t-k+1} \!\! \bold S_{\tau} \big)  \big \|_2  \bigg]{\|\bold x \|}^2    \\ 
     &{\leq}  \rho^{l+k}   \; \| {{\text{diag}(\boldsymbol \phi^{(l)})}} \|_2  \; \| \text{diag}(\boldsymbol \phi^{(k)})\|_2  \;  \;  {\|\bold x \|} ^2    
  \end{split}
  \label{eq_Upsilon}
 \end{equation}
\end{footnotesize}
\vspace{-0.2cm}

\noindent where 
  we  assume  here an upper-bounded  spectral norm of the shift operator 
\textit{i.e.,} $\| \bold S_t  \|_2  \leq \| \bold S  \|_2  \leq \rho$ for all $t$  \cite{GamaIsufiLeus2018,HOPPENMonsalve2019}, implying that:
\vspace{-0.35cm}

\begin{small}
\begin{align}
\begin{split}
\text{tr}(\mathbb{E} \big[\bold y_t {\bold y_t}^{H}])&\leq   \displaystyle\sum_{ {k=0}, {l=0}}^{L}   \rho^{l+k}   \; \| {{\text{diag}(\boldsymbol \phi^{(l)})}} \|_2  \; \| \text{diag}(\boldsymbol \phi^{(k)})\|_2  \;  \;  {\|\bold x \|} ^2    \\ 
&\leq  \;  {\|\bold x \|} ^2  \big ( \rho^{0}   \| {{\text{diag}(\boldsymbol \phi^{(0)})}} \|_2  + \rho^{1}   \| {{\text{diag}(\boldsymbol \phi^{(1)})}} \|_2  +..\\
&+\rho^{L}  \| {{\text{diag}(\boldsymbol \phi^{(L)})}} \|_2   \big)^2  \\ 
\end{split}\label{eq_bound}
\end{align}
\end{small}
\vspace{-0.1cm}

\noindent Observe  that the second term in (\ref{eq_variance})  is  positive i.e., $\text{tr}(\mathbb{E}[\bold y_t]\mathbb{E}[\bold {y_t}]^{H}) {\geq} 0$.
Then,  if we divide  both sides by $N$ in (\ref{eq_bound}) and operate, 
  $\overline{var}  [\bold y_t]$  can be upper bounded by   (\ref{bound_variance}).\qed
  \vspace{-0.25cm}
  
\subsection*{Proof of \textbf{Proposition 2}} 
The disc of radius $R^{\ast}_P=R_P^{\hat{n}_I=1}$ corresponds to  the preventing area 
used by our CDSA protocol when assigning slots in the case that there are at most two  nodes that can transmit simultaneously, which corresponds to one transmitter and one interfering node i.e., $\hat{n}_I=1$.
If all the $N$ sensor nodes are located inside a disc of radius $R^{\ast}_{P}$, this means that 
 $\forall i \in \mathcal{V} \; \text{and}\;  \forall  j \in \mathcal{V}$, 
the distance $d_{i,j} \leq 2\;R^{\ast}_{P}$, which means that
 in the whole network, there is no node $i$ that
can simultaneously transmit with  another node $j$, because our CDSA protocol {cannot}  allocate  the same slot to
these nodes  to transmit simultaneously. This is due to the fact that 
 their preventing areas are overlapping. 
Therefore, each node will be allocated a slot to transmit exclusively alone, making the total number of allocated slots $T_{s}$ equal to $N$.
This implies that   the  number of  slots is $T_{s}=N$ 
  if   and only if there exists a disc of radius $R^{\ast}_{P}$ that contains all the $N$  nodes.
In addition to that, by considering that the $N$ sensor nodes are deployed  uniformly random inside a 2-D  square area of side length $\ell_s$,
  the probability
 that there are any two nodes located at a distance shorter
 than  $2\; R^{\ast}_P$, is simply given
 by:
   \begin{equation}
    \text{prob}_1=\frac{\pi (2 {R^{\ast}_P})^2}{{\ell_s}^2}
   \end{equation}

\noindent This means that the probability
 that there are any two nodes located at a distance larger
 than  $2\;R^{\ast}_P$, is  given by:  
    \begin{equation}
    \text{prob}_2=1-\text{prob}_1=1-\frac{ \pi (2 {R^{\ast}_P})^2 }{{\ell_s}^2}
   \end{equation}
   Therefore,    to  decrease the number of  allocated slots in our CDSA protocol,
the probability of having all  nodes inside a disc of
radius $R^{\ast}_P$ should be decreased, by increasing the probability
 that there are any two nodes located at a distance higher
 than  $2\; R^{\ast}_P$. This can be achieved if the side  of the deployed square area $\ell_s$ is
 selected such that  $\ell_s >> R^{\ast}_P$ and by reducing the value of $\chi$  while still maintainting  the connectivity of the whole network.
 Thus, a small value  of $\chi$ can be selected in the range $\ell_s \sqrt{(\pi N R^2_\text{m})^{-1} \log{N} }<\chi<1$, 
 since for large-scale networks deployed randomly and uniformly, the critical radius for connectivity is $\ell_s \sqrt{(\pi N)^{-1} \log{N} }$,
 as shown in \cite{GuptaKumar2000, AsensioAlonsoLozano2019}.\qed 
\begin{small}
\bibliographystyle{IEEEtran}
 \bibliography{references}
\end{small}
\end{document}